\documentclass[sigconf]{acmart}
\newcommand{\showDOI}[1]{\unskip}
\usepackage{booktabs} 
\usepackage{pgfplots}
\usetikzlibrary{pgfplots.groupplots}
\usepgfplotslibrary{statistics}
\usetikzlibrary{patterns}

\makeatletter
\newcommand\resetstackedplots{
\makeatletter
\pgfplots@stacked@isfirstplottrue
\makeatother
\addplot [forget plot,draw=none] coordinates{(1,0) (2,0) (3,0) (4,0) (5,0) (6,0) (7,0) (8,0) (9,0) (10,0)};
}
\makeatother
\newcommand{\cthree}{$\mathcal{C}3$}
\usepackage{subcaption}

\usepackage{listings}

\definecolor{addcolor}{HTML}{008837}
\definecolor{delcolor}{HTML}{d7191c}
\definecolor{changecolor}{HTML}{2b83ba}

\lstdefinelanguage{diff}[]{java}{
  morekeywords={+,-},
  moredelim=**[l][\color{delcolor}]{-},
  moredelim=**[l][\color{addcolor}]{+},
}
\lstdefinelanguage{java2}[]{java}{
  moredelim=**[il][\color{changecolor}]{?}
}
\usepackage{tikz}
\usepackage{tikz-qtree}
\usetikzlibrary{shapes,fit,calc,positioning,decorations.markings,shadows}
\tikzstyle{iweight}=[outer sep=0pt, inner sep=1pt, fill=black!5, draw=black]

\usepackage{subcaption}

\usepackage{algpseudocode}
\usepackage{algorithm}
\usepackage{setspace}
\usepackage{balance}

\makeatletter
\def\therule{\makebox[\algorithmicindent][l]{}}%

\newtoks\therules
\therules={}
\def\appendto#1#2{\expandafter#1\expandafter{\the#1#2}}
\def\gobblefirst#1{
  #1\expandafter\expandafter\expandafter{\expandafter\@gobble\the#1}}%
\def\LState{\State\unskip\the\therules}
\def\pushindent{\appendto\therules\therule}%
\def\popindent{\gobblefirst\therules}%
\def\printindent{\unskip\the\therules}%
\def\printandpush{\printindent\pushindent}%
\def\popandprint{\popindent\printindent}%

\algdef{SE}[WHILE]{While}{EndWhile}[1]
  {\printandpush\algorithmicwhile\ #1\ \algorithmicdo}
  {\popandprint\algorithmicend\ \algorithmicwhile}%
\algdef{SE}[FOR]{For}{EndFor}[1]
  {\printandpush\algorithmicfor\ #1\ \algorithmicdo}
  {\popandprint\algorithmicend\ \algorithmicfor}%
\algdef{S}[FOR]{ForAll}[1]
  {\printindent\algorithmicforall\ #1\ \algorithmicdo}%
\algdef{SE}[LOOP]{Loop}{EndLoop}
  {\printandpush\algorithmicloop}
  {\popandprint\algorithmicend\ \algorithmicloop}%
\algdef{SE}[REPEAT]{Repeat}{Until}
  {\printandpush\algorithmicrepeat}[1]
  {\popandprint\algorithmicuntil\ #1}%
\algdef{SE}[IF]{If}{EndIf}[1]
  {\printandpush\algorithmicif\ #1\ \algorithmicthen}
  {\popandprint\algorithmicend\ \algorithmicif}%
\algdef{C}[IF]{IF}{ElsIf}[1]
  {\popandprint\pushindent\algorithmicelse\ \algorithmicif\ #1\ \algorithmicthen}%
\algdef{Ce}[ELSE]{IF}{Else}{EndIf}
  {\popandprint\pushindent\algorithmicelse}%
\algdef{SE}[PROCEDURE]{Procedure}{EndProcedure}[2]
   {\printandpush\algorithmicprocedure\ \textproc{#1}\ifthenelse{\equal{#2}{}}{}{(#2)}}%
   {\popandprint\algorithmicend\ \algorithmicprocedure}%
\algdef{SE}[FUNCTION]{Function}{EndFunction}[2]
   {\printandpush\algorithmicfunction\ \textproc{#1}\ifthenelse{\equal{#2}{}}{}{(#2)}}%
   {\popandprint\algorithmicend\ \algorithmicfunction}%

\def\balanceissued{unbalanced}
\let\oldbibitem\bibitem
\def\bibitem{%
  \expandafter\ifx\expandafter\relax\balanceissued\relax\else%
    \balance%
    \gdef\balanceissued{\relax}\fi%
  \oldbibitem}
\makeatother

\algrenewcommand{\alglinenumber}[1]{\ifnum#1<10 \ \fi\number#1:}

   \algtext*{EndWhile}%
   \algtext*{EndFor}%
   \algtext*{EndLoop}%
   \algtext*{EndIf}%
   \algtext*{EndProcedure}%
   \algtext*{EndFunction}%

\def\LEndIf{\EndIf\makeatletter\popindent\makeatother}

\def\LEndWhile{\EndWhile\makeatletter\popindent\makeatother}

\setcopyright{acmlicensed}
\acmDOI{10.1145/3106237.3106276}
\acmISBN{978-1-4503-5105-8/17/09}
\acmConference[ESEC/FSE'17]{2017 11th Joint Meeting of the European Software Engineering Conference and the ACM SIGSOFT Symposium on the Foundations of Software Engineering}{September 4-8, 2017}{Paderborn, Germany} 
\acmYear{2017}
\copyrightyear{2017}
\acmPrice{15.00}

\begin{document}
\title{More Accurate Recommendations for Method-Level Changes}

\author{Georg Dotzler, Marius Kamp, Patrick Kreutzer, Michael Philippsen}
\affiliation{%
  \institution{Programming Systems Group, Friedrich-Alexander University Erlangen-N\"urnberg (FAU), Germany}
}
\email{{georg.dotzler, marius.kamp, patrick.kreutzer, michael.philippsen}@fau.de}

\newcommand{\indexTwoFilesA}{1}
\newcommand{\bugzillaIdTwoFilesA}{77644}
\newcommand{\numberOfMembersTwoFilesA}{4}
\newcommand{\recommendationsAresTwoFilesA}{2}
\newcommand{\foundPatternsAresTwoFilesA}{2}
\newcommand{\precisionAresTwoFilesA}{100}
\newcommand{\recallAresTwoFilesA}{50}
\newcommand{\accuracyTokenAresTwoFilesA}{84}
\newcommand{\accuracyCharactersAresTwoFilesA}{83}
\newcommand{\recommendationsLaseTwoFilesA}{2}
\newcommand{\foundPatternsLaseTwoFilesA}{2}
\newcommand{\precisionLaseTwoFilesA}{100}
\newcommand{\recallLaseTwoFilesA}{50}
\newcommand{\accuracyTokenLaseTwoFilesA}{52}
\newcommand{\accuracyCharactersLaseTwoFilesA}{52}
\newcommand{\indexTwoFilesB}{2}
\newcommand{\bugzillaIdTwoFilesB}{82429}
\newcommand{\numberOfMembersTwoFilesB}{16}
\newcommand{\recommendationsAresTwoFilesB}{21}
\newcommand{\foundPatternsAresTwoFilesB}{14}
\newcommand{\precisionAresTwoFilesB}{67}
\newcommand{\recallAresTwoFilesB}{88}
\newcommand{\accuracyTokenAresMinTwoFilesB}{68}
\newcommand{\accuracyCharactersAresMinTwoFilesB}{76}
\newcommand{\accuracyTokenAresMaxTwoFilesB}{81}
\newcommand{\accuracyCharactersAresMaxTwoFilesB}{84}
\newcommand{\recommendationsLaseTwoFilesB}{14}
\newcommand{\foundPatternsLaseTwoFilesB}{14}
\newcommand{\precisionLaseTwoFilesB}{100}
\newcommand{\recallLaseTwoFilesB}{88}
\newcommand{\accuracyTokenLaseTwoFilesB}{66}
\newcommand{\accuracyCharactersLaseTwoFilesB}{40}
\newcommand{\indexTwoFilesC}{3}
\newcommand{\bugzillaIdTwoFilesC}{114007}
\newcommand{\numberOfMembersTwoFilesC}{4}
\newcommand{\recommendationsAresTwoFilesC}{4}
\newcommand{\foundPatternsAresTwoFilesC}{4}
\newcommand{\precisionAresTwoFilesC}{100}
\newcommand{\recallAresTwoFilesC}{100}
\newcommand{\accuracyTokenAresTwoFilesC}{100}
\newcommand{\accuracyCharactersAresTwoFilesC}{100}
\newcommand{\recommendationsLaseTwoFilesC}{4}
\newcommand{\foundPatternsLaseTwoFilesC}{4}
\newcommand{\precisionLaseTwoFilesC}{100}
\newcommand{\recallLaseTwoFilesC}{100}
\newcommand{\accuracyTokenLaseTwoFilesC}{94}
\newcommand{\accuracyCharactersLaseTwoFilesC}{81}
\newcommand{\indexTwoFilesD}{4}
\newcommand{\bugzillaIdTwoFilesD}{139329$_1$}
\newcommand{\numberOfMembersTwoFilesD}{6}
\newcommand{\recommendationsAresTwoFilesD}{2}
\newcommand{\foundPatternsAresTwoFilesD}{2}
\newcommand{\precisionAresTwoFilesD}{100}
\newcommand{\recallAresTwoFilesD}{33}
\newcommand{\accuracyTokenAresTwoFilesD}{100}
\newcommand{\accuracyCharactersAresTwoFilesD}{100}
\newcommand{\recommendationsLaseTwoFilesD}{2}
\newcommand{\foundPatternsLaseTwoFilesD}{2}
\newcommand{\precisionLaseTwoFilesD}{100}
\newcommand{\recallLaseTwoFilesD}{33}
\newcommand{\accuracyTokenLaseTwoFilesD}{99}
\newcommand{\accuracyCharactersLaseTwoFilesD}{86}
\newcommand{\indexTwoFilesE}{5}
\newcommand{\bugzillaIdTwoFilesE}{142947$_1$}
\newcommand{\numberOfMembersTwoFilesE}{12}
\newcommand{\recommendationsAresTwoFilesE}{12}
\newcommand{\foundPatternsAresTwoFilesE}{12}
\newcommand{\precisionAresTwoFilesE}{100}
\newcommand{\recallAresTwoFilesE}{100}
\newcommand{\accuracyTokenAresTwoFilesE}{100}
\newcommand{\accuracyCharactersAresTwoFilesE}{100}
\newcommand{\recommendationsLaseTwoFilesE}{12}
\newcommand{\foundPatternsLaseTwoFilesE}{12}
\newcommand{\precisionLaseTwoFilesE}{100}
\newcommand{\recallLaseTwoFilesE}{100}
\newcommand{\accuracyTokenLaseTwoFilesE}{98}
\newcommand{\accuracyCharactersLaseTwoFilesE}{89}
\newcommand{\indexTwoFilesF}{6}
\newcommand{\bugzillaIdTwoFilesF}{91937}
\newcommand{\numberOfMembersTwoFilesF}{3}
\newcommand{\recommendationsAresTwoFilesF}{3}
\newcommand{\foundPatternsAresTwoFilesF}{3}
\newcommand{\precisionAresTwoFilesF}{100}
\newcommand{\recallAresTwoFilesF}{100}
\newcommand{\accuracyTokenAresTwoFilesF}{100}
\newcommand{\accuracyCharactersAresTwoFilesF}{100}
\newcommand{\recommendationsLaseTwoFilesF}{3}
\newcommand{\foundPatternsLaseTwoFilesF}{3}
\newcommand{\precisionLaseTwoFilesF}{100}
\newcommand{\recallLaseTwoFilesF}{100}
\newcommand{\accuracyTokenLaseTwoFilesF}{50}
\newcommand{\accuracyCharactersLaseTwoFilesF}{37}
\newcommand{\indexTwoFilesG}{7}
\newcommand{\bugzillaIdTwoFilesG}{103863}
\newcommand{\numberOfMembersTwoFilesG}{7}
\newcommand{\recommendationsAresTwoFilesG}{7}
\newcommand{\foundPatternsAresTwoFilesG}{7}
\newcommand{\precisionAresTwoFilesG}{100}
\newcommand{\recallAresTwoFilesG}{100}
\newcommand{\accuracyTokenAresTwoFilesG}{100}
\newcommand{\accuracyCharactersAresTwoFilesG}{100}
\newcommand{\recommendationsLaseTwoFilesG}{7}
\newcommand{\foundPatternsLaseTwoFilesG}{7}
\newcommand{\precisionLaseTwoFilesG}{100}
\newcommand{\recallLaseTwoFilesG}{100}
\newcommand{\accuracyTokenLaseTwoFilesG}{32}
\newcommand{\accuracyCharactersLaseTwoFilesG}{43}
\newcommand{\indexTwoFilesH}{8}
\newcommand{\bugzillaIdTwoFilesH}{129314}
\newcommand{\numberOfMembersTwoFilesH}{4}
\newcommand{\recommendationsAresTwoFilesH}{3}
\newcommand{\foundPatternsAresTwoFilesH}{2}
\newcommand{\precisionAresTwoFilesH}{67}
\newcommand{\recallAresTwoFilesH}{50}
\newcommand{\accuracyTokenAresMinTwoFilesH}{94}
\newcommand{\accuracyCharactersAresMinTwoFilesH}{94}
\newcommand{\accuracyTokenAresMaxTwoFilesH}{100}
\newcommand{\accuracyCharactersAresMaxTwoFilesH}{100}
\newcommand{\recommendationsLaseTwoFilesH}{4}
\newcommand{\foundPatternsLaseTwoFilesH}{4}
\newcommand{\precisionLaseTwoFilesH}{100}
\newcommand{\recallLaseTwoFilesH}{100}
\newcommand{\accuracyTokenLaseTwoFilesH}{96}
\newcommand{\accuracyCharactersLaseTwoFilesH}{80}
\newcommand{\indexTwoFilesI}{9}
\newcommand{\bugzillaIdTwoFilesI}{134091}
\newcommand{\numberOfMembersTwoFilesI}{4}
\newcommand{\recommendationsAresTwoFilesI}{4}
\newcommand{\foundPatternsAresTwoFilesI}{4}
\newcommand{\precisionAresTwoFilesI}{100}
\newcommand{\recallAresTwoFilesI}{100}
\newcommand{\accuracyTokenAresTwoFilesI}{99}
\newcommand{\accuracyCharactersAresTwoFilesI}{99}
\newcommand{\recommendationsLaseTwoFilesI}{4}
\newcommand{\foundPatternsLaseTwoFilesI}{4}
\newcommand{\precisionLaseTwoFilesI}{100}
\newcommand{\recallLaseTwoFilesI}{100}
\newcommand{\accuracyTokenLaseTwoFilesI}{79}
\newcommand{\accuracyCharactersLaseTwoFilesI}{56}
\newcommand{\indexTwoFilesJ}{10}
\newcommand{\bugzillaIdTwoFilesJ}{139329$_2$}
\newcommand{\numberOfMembersTwoFilesJ}{3}
\newcommand{\recommendationsAresTwoFilesJ}{3}
\newcommand{\foundPatternsAresTwoFilesJ}{3}
\newcommand{\precisionAresTwoFilesJ}{100}
\newcommand{\recallAresTwoFilesJ}{100}
\newcommand{\accuracyTokenAresTwoFilesJ}{95}
\newcommand{\accuracyCharactersAresTwoFilesJ}{94}
\newcommand{\recommendationsLaseTwoFilesJ}{4}
\newcommand{\foundPatternsLaseTwoFilesJ}{3}
\newcommand{\precisionLaseTwoFilesJ}{75}
\newcommand{\recallLaseTwoFilesJ}{100}
\newcommand{\accuracyTokenLaseTwoFilesJ}{63}
\newcommand{\accuracyCharactersLaseTwoFilesJ}{64}
\newcommand{\indexTwoFilesK}{11}
\newcommand{\bugzillaIdTwoFilesK}{139329$_3$}
\newcommand{\numberOfMembersTwoFilesK}{3}
\newcommand{\recommendationsAresTwoFilesK}{3}
\newcommand{\foundPatternsAresTwoFilesK}{3}
\newcommand{\precisionAresTwoFilesK}{100}
\newcommand{\recallAresTwoFilesK}{100}
\newcommand{\accuracyTokenAresTwoFilesK}{100}
\newcommand{\accuracyCharactersAresTwoFilesK}{100}
\newcommand{\recommendationsLaseTwoFilesK}{3}
\newcommand{\foundPatternsLaseTwoFilesK}{3}
\newcommand{\precisionLaseTwoFilesK}{100}
\newcommand{\recallLaseTwoFilesK}{100}
\newcommand{\accuracyTokenLaseTwoFilesK}{92}
\newcommand{\accuracyCharactersLaseTwoFilesK}{75}
\newcommand{\indexTwoFilesL}{12}
\newcommand{\bugzillaIdTwoFilesL}{142947$_2$}
\newcommand{\numberOfMembersTwoFilesL}{9}
\newcommand{\recommendationsAresTwoFilesL}{7}
\newcommand{\foundPatternsAresTwoFilesL}{6}
\newcommand{\precisionAresTwoFilesL}{86}
\newcommand{\recallAresTwoFilesL}{67}
\newcommand{\accuracyTokenAresMinTwoFilesL}{78}
\newcommand{\accuracyCharactersAresMinTwoFilesL}{74}
\newcommand{\accuracyTokenAresMaxTwoFilesL}{89}
\newcommand{\accuracyCharactersAresMaxTwoFilesL}{87}
\newcommand{\recommendationsLaseTwoFilesL}{12}
\newcommand{\foundPatternsLaseTwoFilesL}{9}
\newcommand{\precisionLaseTwoFilesL}{75}
\newcommand{\recallLaseTwoFilesL}{100}
\newcommand{\accuracyTokenLaseTwoFilesL}{74}
\newcommand{\accuracyCharactersLaseTwoFilesL}{61}
\newcommand{\indexTwoFilesM}{13}
\newcommand{\bugzillaIdTwoFilesM}{76182}
\newcommand{\numberOfMembersTwoFilesM}{6}
\newcommand{\recommendationsAresTwoFilesM}{6}
\newcommand{\foundPatternsAresTwoFilesM}{6}
\newcommand{\precisionAresTwoFilesM}{100}
\newcommand{\recallAresTwoFilesM}{100}
\newcommand{\accuracyTokenAresTwoFilesM}{92}
\newcommand{\accuracyCharactersAresTwoFilesM}{92}
\newcommand{\recommendationsLaseTwoFilesM}{6}
\newcommand{\foundPatternsLaseTwoFilesM}{6}
\newcommand{\precisionLaseTwoFilesM}{100}
\newcommand{\recallLaseTwoFilesM}{100}
\newcommand{\accuracyTokenLaseTwoFilesM}{56}
\newcommand{\accuracyCharactersLaseTwoFilesM}{58}
\newcommand{\indexTwoFilesN}{14}
\newcommand{\bugzillaIdTwoFilesN}{77194}
\newcommand{\numberOfMembersTwoFilesN}{3}
\newcommand{\recommendationsAresTwoFilesN}{2}
\newcommand{\foundPatternsAresTwoFilesN}{2}
\newcommand{\precisionAresTwoFilesN}{100}
\newcommand{\recallAresTwoFilesN}{67}
\newcommand{\accuracyTokenAresTwoFilesN}{98}
\newcommand{\accuracyCharactersAresTwoFilesN}{98}
\newcommand{\recommendationsLaseTwoFilesN}{3}
\newcommand{\foundPatternsLaseTwoFilesN}{3}
\newcommand{\precisionLaseTwoFilesN}{100}
\newcommand{\recallLaseTwoFilesN}{100}
\newcommand{\accuracyTokenLaseTwoFilesN}{79}
\newcommand{\accuracyCharactersLaseTwoFilesN}{58}
\newcommand{\indexTwoFilesO}{15}
\newcommand{\bugzillaIdTwoFilesO}{86079$_1$}
\newcommand{\numberOfMembersTwoFilesO}{3}
\newcommand{\recommendationsAresTwoFilesO}{2}
\newcommand{\foundPatternsAresTwoFilesO}{2}
\newcommand{\precisionAresTwoFilesO}{100}
\newcommand{\recallAresTwoFilesO}{67}
\newcommand{\accuracyTokenAresTwoFilesO}{100}
\newcommand{\accuracyCharactersAresTwoFilesO}{100}
\newcommand{\recommendationsLaseTwoFilesO}{3}
\newcommand{\foundPatternsLaseTwoFilesO}{3}
\newcommand{\precisionLaseTwoFilesO}{100}
\newcommand{\recallLaseTwoFilesO}{100}
\newcommand{\accuracyTokenLaseTwoFilesO}{92}
\newcommand{\accuracyCharactersLaseTwoFilesO}{69}
\newcommand{\indexTwoFilesP}{16}
\newcommand{\bugzillaIdTwoFilesP}{95409}
\newcommand{\numberOfMembersTwoFilesP}{9}
\newcommand{\recommendationsAresTwoFilesP}{4}
\newcommand{\foundPatternsAresTwoFilesP}{4}
\newcommand{\precisionAresTwoFilesP}{100}
\newcommand{\recallAresTwoFilesP}{44}
\newcommand{\accuracyTokenAresMinTwoFilesP}{42}
\newcommand{\accuracyCharactersAresMinTwoFilesP}{43}
\newcommand{\accuracyTokenAresMaxTwoFilesP}{85}
\newcommand{\accuracyCharactersAresMaxTwoFilesP}{84}
\newcommand{\recommendationsLaseTwoFilesP}{8}
\newcommand{\foundPatternsLaseTwoFilesP}{8}
\newcommand{\precisionLaseTwoFilesP}{100}
\newcommand{\recallLaseTwoFilesP}{89}
\newcommand{\accuracyTokenLaseTwoFilesP}{49}
\newcommand{\accuracyCharactersLaseTwoFilesP}{46}
\newcommand{\indexTwoFilesQ}{17}
\newcommand{\bugzillaIdTwoFilesQ}{97981}
\newcommand{\numberOfMembersTwoFilesQ}{4}
\newcommand{\recommendationsAresTwoFilesQ}{3}
\newcommand{\foundPatternsAresTwoFilesQ}{3}
\newcommand{\precisionAresTwoFilesQ}{100}
\newcommand{\recallAresTwoFilesQ}{75}
\newcommand{\accuracyTokenAresTwoFilesQ}{100}
\newcommand{\accuracyCharactersAresTwoFilesQ}{100}
\newcommand{\recommendationsLaseTwoFilesQ}{3}
\newcommand{\foundPatternsLaseTwoFilesQ}{3}
\newcommand{\precisionLaseTwoFilesQ}{100}
\newcommand{\recallLaseTwoFilesQ}{75}
\newcommand{\accuracyTokenLaseTwoFilesQ}{71}
\newcommand{\accuracyCharactersLaseTwoFilesQ}{56}
\newcommand{\indexTwoFilesR}{18}
\newcommand{\bugzillaIdTwoFilesR}{76391}
\newcommand{\numberOfMembersTwoFilesR}{6}
\newcommand{\recommendationsAresTwoFilesR}{3}
\newcommand{\foundPatternsAresTwoFilesR}{3}
\newcommand{\precisionAresTwoFilesR}{100}
\newcommand{\recallAresTwoFilesR}{50}
\newcommand{\accuracyTokenAresTwoFilesR}{100}
\newcommand{\accuracyCharactersAresTwoFilesR}{100}
\newcommand{\recommendationsLaseTwoFilesR}{3}
\newcommand{\foundPatternsLaseTwoFilesR}{3}
\newcommand{\precisionLaseTwoFilesR}{100}
\newcommand{\recallLaseTwoFilesR}{50}
\newcommand{\accuracyTokenLaseTwoFilesR}{97}
\newcommand{\accuracyCharactersLaseTwoFilesR}{86}
\newcommand{\indexTwoFilesS}{19}
\newcommand{\bugzillaIdTwoFilesS}{89785}
\newcommand{\numberOfMembersTwoFilesS}{5}
\newcommand{\recommendationsAresTwoFilesS}{5}
\newcommand{\foundPatternsAresTwoFilesS}{5}
\newcommand{\precisionAresTwoFilesS}{100}
\newcommand{\recallAresTwoFilesS}{100}
\newcommand{\accuracyTokenAresTwoFilesS}{97}
\newcommand{\accuracyCharactersAresTwoFilesS}{94}
\newcommand{\recommendationsLaseTwoFilesS}{5}
\newcommand{\foundPatternsLaseTwoFilesS}{5}
\newcommand{\precisionLaseTwoFilesS}{100}
\newcommand{\recallLaseTwoFilesS}{100}
\newcommand{\accuracyTokenLaseTwoFilesS}{91}
\newcommand{\accuracyCharactersLaseTwoFilesS}{75}
\newcommand{\indexTwoFilesT}{20}
\newcommand{\bugzillaIdTwoFilesT}{79107}
\newcommand{\numberOfMembersTwoFilesT}{10}
\newcommand{\recommendationsAresTwoFilesT}{12}
\newcommand{\foundPatternsAresTwoFilesT}{4}
\newcommand{\precisionAresTwoFilesT}{33}
\newcommand{\recallAresTwoFilesT}{40}
\newcommand{\accuracyTokenAresTwoFilesT}{99}
\newcommand{\accuracyCharactersAresTwoFilesT}{99}
\newcommand{\recommendationsLaseTwoFilesT}{26}
\newcommand{\foundPatternsLaseTwoFilesT}{10}
\newcommand{\precisionLaseTwoFilesT}{38}
\newcommand{\recallLaseTwoFilesT}{100}
\newcommand{\accuracyTokenLaseTwoFilesT}{96}
\newcommand{\accuracyCharactersLaseTwoFilesT}{75}
\newcommand{\indexTwoFilesU}{21}
\newcommand{\bugzillaIdTwoFilesU}{86079$_2$}
\newcommand{\numberOfMembersTwoFilesU}{3}
\newcommand{\recommendationsAresTwoFilesU}{2}
\newcommand{\foundPatternsAresTwoFilesU}{2}
\newcommand{\precisionAresTwoFilesU}{100}
\newcommand{\recallAresTwoFilesU}{67}
\newcommand{\accuracyTokenAresTwoFilesU}{100}
\newcommand{\accuracyCharactersAresTwoFilesU}{100}
\newcommand{\recommendationsLaseTwoFilesU}{2}
\newcommand{\foundPatternsLaseTwoFilesU}{2}
\newcommand{\precisionLaseTwoFilesU}{100}
\newcommand{\recallLaseTwoFilesU}{67}
\newcommand{\accuracyTokenLaseTwoFilesU}{99}
\newcommand{\accuracyCharactersLaseTwoFilesU}{73}
\newcommand{\indexTwoFilesV}{22}
\newcommand{\bugzillaIdTwoFilesV}{95116}
\newcommand{\numberOfMembersTwoFilesV}{5}
\newcommand{\recommendationsAresTwoFilesV}{4}
\newcommand{\foundPatternsAresTwoFilesV}{4}
\newcommand{\precisionAresTwoFilesV}{100}
\newcommand{\recallAresTwoFilesV}{80}
\newcommand{\accuracyTokenAresTwoFilesV}{100}
\newcommand{\accuracyCharactersAresTwoFilesV}{100}
\newcommand{\recommendationsLaseTwoFilesV}{4}
\newcommand{\foundPatternsLaseTwoFilesV}{4}
\newcommand{\precisionLaseTwoFilesV}{100}
\newcommand{\recallLaseTwoFilesV}{80}
\newcommand{\accuracyTokenLaseTwoFilesV}{97}
\newcommand{\accuracyCharactersLaseTwoFilesV}{83}
\newcommand{\indexTwoFilesW}{23}
\newcommand{\bugzillaIdTwoFilesW}{98198}
\newcommand{\numberOfMembersTwoFilesW}{15}
\newcommand{\recommendationsAresTwoFilesW}{38}
\newcommand{\foundPatternsAresTwoFilesW}{10}
\newcommand{\precisionAresTwoFilesW}{26}
\newcommand{\recallAresTwoFilesW}{67}
\newcommand{\accuracyTokenAresMinTwoFilesW}{71}
\newcommand{\accuracyCharactersAresMinTwoFilesW}{75}
\newcommand{\accuracyTokenAresMaxTwoFilesW}{93}
\newcommand{\accuracyCharactersAresMaxTwoFilesW}{94}
\newcommand{\recommendationsLaseTwoFilesW}{67}
\newcommand{\foundPatternsLaseTwoFilesW}{12}
\newcommand{\precisionLaseTwoFilesW}{18}
\newcommand{\recallLaseTwoFilesW}{80}
\newcommand{\accuracyTokenLaseTwoFilesW}{61}
\newcommand{\accuracyCharactersLaseTwoFilesW}{51}
\newcommand{\indexAllA}{1}
\newcommand{\bugzillaIdAllA}{77644}
\newcommand{\numberOfMembersAllA}{4}
\newcommand{\recommendationsAresAllA}{4}
\newcommand{\foundPatternsAresAllA}{4}
\newcommand{\precisionAresAllA}{100}
\newcommand{\recallAresAllA}{100}
\newcommand{\accuracyTokenAresAllA}{90}
\newcommand{\accuracyCharactersAresAllA}{89}
\newcommand{\recommendationsLaseAllA}{0}
\newcommand{\foundPatternsLaseAllA}{0}
\newcommand{\precisionLaseAllA}{0}
\newcommand{\recallLaseAllA}{0}
\newcommand{\accuracyTokenLaseAllA}{-100}
\newcommand{\accuracyCharactersLaseAllA}{-100}
\newcommand{\indexAllB}{2}
\newcommand{\bugzillaIdAllB}{82429}
\newcommand{\numberOfMembersAllB}{16}
\newcommand{\recommendationsAresAllB}{25}
\newcommand{\foundPatternsAresAllB}{16}
\newcommand{\precisionAresAllB}{64}
\newcommand{\recallAresAllB}{100}
\newcommand{\accuracyTokenAresAllB}{54}
\newcommand{\accuracyCharactersAresAllB}{61}
\newcommand{\recommendationsLaseAllB}{0}
\newcommand{\foundPatternsLaseAllB}{0}
\newcommand{\precisionLaseAllB}{0}
\newcommand{\recallLaseAllB}{0}
\newcommand{\accuracyTokenLaseAllB}{-100}
\newcommand{\accuracyCharactersLaseAllB}{-100}
\newcommand{\indexAllC}{3}
\newcommand{\bugzillaIdAllC}{114007}
\newcommand{\numberOfMembersAllC}{4}
\newcommand{\recommendationsAresAllC}{4}
\newcommand{\foundPatternsAresAllC}{4}
\newcommand{\precisionAresAllC}{100}
\newcommand{\recallAresAllC}{100}
\newcommand{\accuracyTokenAresAllC}{100}
\newcommand{\accuracyCharactersAresAllC}{100}
\newcommand{\recommendationsLaseAllC}{4}
\newcommand{\foundPatternsLaseAllC}{4}
\newcommand{\precisionLaseAllC}{100}
\newcommand{\recallLaseAllC}{100}
\newcommand{\accuracyTokenLaseAllC}{94}
\newcommand{\accuracyCharactersLaseAllC}{81}
\newcommand{\indexAllD}{4}
\newcommand{\bugzillaIdAllD}{139329$_1$}
\newcommand{\numberOfMembersAllD}{6}
\newcommand{\recommendationsAresAllD}{35}
\newcommand{\foundPatternsAresAllD}{6}
\newcommand{\precisionAresAllD}{17}
\newcommand{\recallAresAllD}{100}
\newcommand{\accuracyTokenAresAllD}{80}
\newcommand{\accuracyCharactersAresAllD}{82}
\newcommand{\recommendationsLaseAllD}{6}
\newcommand{\foundPatternsLaseAllD}{6}
\newcommand{\precisionLaseAllD}{100}
\newcommand{\recallLaseAllD}{100}
\newcommand{\accuracyTokenLaseAllD}{78}
\newcommand{\accuracyCharactersLaseAllD}{64}
\newcommand{\indexAllE}{5}
\newcommand{\bugzillaIdAllE}{142947$_1$}
\newcommand{\numberOfMembersAllE}{12}
\newcommand{\recommendationsAresAllE}{12}
\newcommand{\foundPatternsAresAllE}{12}
\newcommand{\precisionAresAllE}{100}
\newcommand{\recallAresAllE}{100}
\newcommand{\accuracyTokenAresAllE}{100}
\newcommand{\accuracyCharactersAresAllE}{100}
\newcommand{\recommendationsLaseAllE}{6}
\newcommand{\foundPatternsLaseAllE}{6}
\newcommand{\precisionLaseAllE}{100}
\newcommand{\recallLaseAllE}{50}
\newcommand{\accuracyTokenLaseAllE}{98}
\newcommand{\accuracyCharactersLaseAllE}{88}
\newcommand{\indexAllF}{6}
\newcommand{\bugzillaIdAllF}{91937}
\newcommand{\numberOfMembersAllF}{3}
\newcommand{\recommendationsAresAllF}{3}
\newcommand{\foundPatternsAresAllF}{3}
\newcommand{\precisionAresAllF}{100}
\newcommand{\recallAresAllF}{100}
\newcommand{\accuracyTokenAresAllF}{100}
\newcommand{\accuracyCharactersAresAllF}{100}
\newcommand{\recommendationsLaseAllF}{3}
\newcommand{\foundPatternsLaseAllF}{3}
\newcommand{\precisionLaseAllF}{100}
\newcommand{\recallLaseAllF}{100}
\newcommand{\accuracyTokenLaseAllF}{94}
\newcommand{\accuracyCharactersLaseAllF}{70}
\newcommand{\indexAllG}{7}
\newcommand{\bugzillaIdAllG}{103863}
\newcommand{\numberOfMembersAllG}{7}
\newcommand{\recommendationsAresAllG}{7}
\newcommand{\foundPatternsAresAllG}{7}
\newcommand{\precisionAresAllG}{100}
\newcommand{\recallAresAllG}{100}
\newcommand{\accuracyTokenAresAllG}{100}
\newcommand{\accuracyCharactersAresAllG}{100}
\newcommand{\recommendationsLaseAllG}{5}
\newcommand{\foundPatternsLaseAllG}{5}
\newcommand{\precisionLaseAllG}{100}
\newcommand{\recallLaseAllG}{71}
\newcommand{\accuracyTokenLaseAllG}{32}
\newcommand{\accuracyCharactersLaseAllG}{43}
\newcommand{\indexAllH}{8}
\newcommand{\bugzillaIdAllH}{129314}
\newcommand{\numberOfMembersAllH}{4}
\newcommand{\recommendationsAresAllH}{5}
\newcommand{\foundPatternsAresAllH}{4}
\newcommand{\precisionAresAllH}{80}
\newcommand{\recallAresAllH}{100}
\newcommand{\accuracyTokenAresMinAllH}{95}
\newcommand{\accuracyCharactersAresMinAllH}{95}
\newcommand{\accuracyTokenAresMaxAllH}{100}
\newcommand{\accuracyCharactersAresMaxAllH}{100}
\newcommand{\recommendationsLaseAllH}{4}
\newcommand{\foundPatternsLaseAllH}{4}
\newcommand{\precisionLaseAllH}{100}
\newcommand{\recallLaseAllH}{100}
\newcommand{\accuracyTokenLaseAllH}{87}
\newcommand{\accuracyCharactersLaseAllH}{71}
\newcommand{\indexAllI}{9}
\newcommand{\bugzillaIdAllI}{134091}
\newcommand{\numberOfMembersAllI}{4}
\newcommand{\recommendationsAresAllI}{4}
\newcommand{\foundPatternsAresAllI}{4}
\newcommand{\precisionAresAllI}{100}
\newcommand{\recallAresAllI}{100}
\newcommand{\accuracyTokenAresAllI}{99}
\newcommand{\accuracyCharactersAresAllI}{99}
\newcommand{\recommendationsLaseAllI}{0}
\newcommand{\foundPatternsLaseAllI}{0}
\newcommand{\precisionLaseAllI}{0}
\newcommand{\recallLaseAllI}{0}
\newcommand{\accuracyTokenLaseAllI}{-100}
\newcommand{\accuracyCharactersLaseAllI}{-100}
\newcommand{\indexAllJ}{10}
\newcommand{\bugzillaIdAllJ}{139329$_2$}
\newcommand{\numberOfMembersAllJ}{3}
\newcommand{\recommendationsAresAllJ}{3}
\newcommand{\foundPatternsAresAllJ}{3}
\newcommand{\precisionAresAllJ}{100}
\newcommand{\recallAresAllJ}{100}
\newcommand{\accuracyTokenAresMinAllJ}{57}
\newcommand{\accuracyCharactersAresMinAllJ}{56}
\newcommand{\accuracyTokenAresMaxAllJ}{100}
\newcommand{\accuracyCharactersAresMaxAllJ}{100}
\newcommand{\recommendationsLaseAllJ}{3}
\newcommand{\foundPatternsLaseAllJ}{3}
\newcommand{\precisionLaseAllJ}{100}
\newcommand{\recallLaseAllJ}{100}
\newcommand{\accuracyTokenLaseAllJ}{47}
\newcommand{\accuracyCharactersLaseAllJ}{43}
\newcommand{\indexAllK}{11}
\newcommand{\bugzillaIdAllK}{139329$_3$}
\newcommand{\numberOfMembersAllK}{3}
\newcommand{\recommendationsAresAllK}{3}
\newcommand{\foundPatternsAresAllK}{3}
\newcommand{\precisionAresAllK}{100}
\newcommand{\recallAresAllK}{100}
\newcommand{\accuracyTokenAresAllK}{100}
\newcommand{\accuracyCharactersAresAllK}{100}
\newcommand{\recommendationsLaseAllK}{4}
\newcommand{\foundPatternsLaseAllK}{3}
\newcommand{\precisionLaseAllK}{75}
\newcommand{\recallLaseAllK}{100}
\newcommand{\accuracyTokenLaseAllK}{63}
\newcommand{\accuracyCharactersLaseAllK}{64}
\newcommand{\indexAllL}{12}
\newcommand{\bugzillaIdAllL}{142947$_2$}
\newcommand{\numberOfMembersAllL}{9}
\newcommand{\recommendationsAresAllL}{12}
\newcommand{\foundPatternsAresAllL}{9}
\newcommand{\precisionAresAllL}{75}
\newcommand{\recallAresAllL}{100}
\newcommand{\accuracyTokenAresMinAllL}{78}
\newcommand{\accuracyCharactersAresMinAllL}{75}
\newcommand{\accuracyTokenAresMaxAllL}{90}
\newcommand{\accuracyCharactersAresMaxAllL}{89}
\newcommand{\recommendationsLaseAllL}{9}
\newcommand{\foundPatternsLaseAllL}{6}
\newcommand{\precisionLaseAllL}{67}
\newcommand{\recallLaseAllL}{67}
\newcommand{\accuracyTokenLaseAllL}{70}
\newcommand{\accuracyCharactersLaseAllL}{52}
\newcommand{\indexAllM}{13}
\newcommand{\bugzillaIdAllM}{76182}
\newcommand{\numberOfMembersAllM}{6}
\newcommand{\recommendationsAresAllM}{6}
\newcommand{\foundPatternsAresAllM}{6}
\newcommand{\precisionAresAllM}{100}
\newcommand{\recallAresAllM}{100}
\newcommand{\accuracyTokenAresAllM}{92}
\newcommand{\accuracyCharactersAresAllM}{92}
\newcommand{\recommendationsLaseAllM}{6}
\newcommand{\foundPatternsLaseAllM}{6}
\newcommand{\precisionLaseAllM}{100}
\newcommand{\recallLaseAllM}{100}
\newcommand{\accuracyTokenLaseAllM}{56}
\newcommand{\accuracyCharactersLaseAllM}{58}
\newcommand{\indexAllN}{14}
\newcommand{\bugzillaIdAllN}{77194}
\newcommand{\numberOfMembersAllN}{3}
\newcommand{\recommendationsAresAllN}{3}
\newcommand{\foundPatternsAresAllN}{3}
\newcommand{\precisionAresAllN}{100}
\newcommand{\recallAresAllN}{100}
\newcommand{\accuracyTokenAresAllN}{96}
\newcommand{\accuracyCharactersAresAllN}{97}
\newcommand{\recommendationsLaseAllN}{0}
\newcommand{\foundPatternsLaseAllN}{0}
\newcommand{\precisionLaseAllN}{0}
\newcommand{\recallLaseAllN}{0}
\newcommand{\accuracyTokenLaseAllN}{-100}
\newcommand{\accuracyCharactersLaseAllN}{-100}
\newcommand{\indexAllO}{15}
\newcommand{\bugzillaIdAllO}{86079$_1$}
\newcommand{\numberOfMembersAllO}{3}
\newcommand{\recommendationsAresAllO}{3}
\newcommand{\foundPatternsAresAllO}{3}
\newcommand{\precisionAresAllO}{100}
\newcommand{\recallAresAllO}{100}
\newcommand{\accuracyTokenAresAllO}{93}
\newcommand{\accuracyCharactersAresAllO}{92}
\newcommand{\recommendationsLaseAllO}{3}
\newcommand{\foundPatternsLaseAllO}{3}
\newcommand{\precisionLaseAllO}{100}
\newcommand{\recallLaseAllO}{100}
\newcommand{\accuracyTokenLaseAllO}{50}
\newcommand{\accuracyCharactersLaseAllO}{40}
\newcommand{\indexAllP}{16}
\newcommand{\bugzillaIdAllP}{95409}
\newcommand{\numberOfMembersAllP}{9}
\newcommand{\recommendationsAresAllP}{18}
\newcommand{\foundPatternsAresAllP}{9}
\newcommand{\precisionAresAllP}{50}
\newcommand{\recallAresAllP}{100}
\newcommand{\accuracyTokenAresMinAllP}{45}
\newcommand{\accuracyCharactersAresMinAllP}{47}
\newcommand{\accuracyTokenAresMaxAllP}{68}
\newcommand{\accuracyCharactersAresMaxAllP}{69}
\newcommand{\recommendationsLaseAllP}{8}
\newcommand{\foundPatternsLaseAllP}{8}
\newcommand{\precisionLaseAllP}{100}
\newcommand{\recallLaseAllP}{89}
\newcommand{\accuracyTokenLaseAllP}{49}
\newcommand{\accuracyCharactersLaseAllP}{46}
\newcommand{\indexAllQ}{17}
\newcommand{\bugzillaIdAllQ}{97981}
\newcommand{\numberOfMembersAllQ}{4}
\newcommand{\recommendationsAresAllQ}{4}
\newcommand{\foundPatternsAresAllQ}{4}
\newcommand{\precisionAresAllQ}{100}
\newcommand{\recallAresAllQ}{100}
\newcommand{\accuracyTokenAresAllQ}{100}
\newcommand{\accuracyCharactersAresAllQ}{100}
\newcommand{\recommendationsLaseAllQ}{4}
\newcommand{\foundPatternsLaseAllQ}{4}
\newcommand{\precisionLaseAllQ}{100}
\newcommand{\recallLaseAllQ}{100}
\newcommand{\accuracyTokenLaseAllQ}{75}
\newcommand{\accuracyCharactersLaseAllQ}{60}
\newcommand{\indexAllR}{18}
\newcommand{\bugzillaIdAllR}{76391}
\newcommand{\numberOfMembersAllR}{6}
\newcommand{\recommendationsAresAllR}{9}
\newcommand{\foundPatternsAresAllR}{6}
\newcommand{\precisionAresAllR}{67}
\newcommand{\recallAresAllR}{100}
\newcommand{\accuracyTokenAresAllR}{100}
\newcommand{\accuracyCharactersAresAllR}{100}
\newcommand{\recommendationsLaseAllR}{6}
\newcommand{\foundPatternsLaseAllR}{6}
\newcommand{\precisionLaseAllR}{100}
\newcommand{\recallLaseAllR}{100}
\newcommand{\accuracyTokenLaseAllR}{96}
\newcommand{\accuracyCharactersLaseAllR}{85}
\newcommand{\indexAllS}{19}
\newcommand{\bugzillaIdAllS}{89785}
\newcommand{\numberOfMembersAllS}{5}
\newcommand{\recommendationsAresAllS}{5}
\newcommand{\foundPatternsAresAllS}{5}
\newcommand{\precisionAresAllS}{100}
\newcommand{\recallAresAllS}{100}
\newcommand{\accuracyTokenAresMinAllS}{84}
\newcommand{\accuracyCharactersAresMinAllS}{79}
\newcommand{\accuracyTokenAresMaxAllS}{100}
\newcommand{\accuracyCharactersAresMaxAllS}{99}
\newcommand{\recommendationsLaseAllS}{5}
\newcommand{\foundPatternsLaseAllS}{5}
\newcommand{\precisionLaseAllS}{100}
\newcommand{\recallLaseAllS}{100}
\newcommand{\accuracyTokenLaseAllS}{91}
\newcommand{\accuracyCharactersLaseAllS}{75}
\newcommand{\indexAllT}{20}
\newcommand{\bugzillaIdAllT}{79107}
\newcommand{\numberOfMembersAllT}{10}
\newcommand{\recommendationsAresAllT}{27}
\newcommand{\foundPatternsAresAllT}{10}
\newcommand{\precisionAresAllT}{37}
\newcommand{\recallAresAllT}{100}
\newcommand{\accuracyTokenAresAllT}{98}
\newcommand{\accuracyCharactersAresAllT}{97}
\newcommand{\recommendationsLaseAllT}{0}
\newcommand{\foundPatternsLaseAllT}{0}
\newcommand{\precisionLaseAllT}{0}
\newcommand{\recallLaseAllT}{0}
\newcommand{\accuracyTokenLaseAllT}{-100}
\newcommand{\accuracyCharactersLaseAllT}{-100}
\newcommand{\indexAllU}{21}
\newcommand{\bugzillaIdAllU}{86079$_2$}
\newcommand{\numberOfMembersAllU}{3}
\newcommand{\recommendationsAresAllU}{3}
\newcommand{\foundPatternsAresAllU}{3}
\newcommand{\precisionAresAllU}{100}
\newcommand{\recallAresAllU}{100}
\newcommand{\accuracyTokenAresMinAllU}{73}
\newcommand{\accuracyCharactersAresMinAllU}{75}
\newcommand{\accuracyTokenAresMaxAllU}{100}
\newcommand{\accuracyCharactersAresMaxAllU}{100}
\newcommand{\recommendationsLaseAllU}{2}
\newcommand{\foundPatternsLaseAllU}{2}
\newcommand{\precisionLaseAllU}{100}
\newcommand{\recallLaseAllU}{67}
\newcommand{\accuracyTokenLaseAllU}{99}
\newcommand{\accuracyCharactersLaseAllU}{73}
\newcommand{\indexAllV}{22}
\newcommand{\bugzillaIdAllV}{95116}
\newcommand{\numberOfMembersAllV}{5}
\newcommand{\recommendationsAresAllV}{5}
\newcommand{\foundPatternsAresAllV}{5}
\newcommand{\precisionAresAllV}{100}
\newcommand{\recallAresAllV}{100}
\newcommand{\accuracyTokenAresAllV}{100}
\newcommand{\accuracyCharactersAresAllV}{100}
\newcommand{\recommendationsLaseAllV}{5}
\newcommand{\foundPatternsLaseAllV}{5}
\newcommand{\precisionLaseAllV}{100}
\newcommand{\recallLaseAllV}{100}
\newcommand{\accuracyTokenLaseAllV}{97}
\newcommand{\accuracyCharactersLaseAllV}{83}
\newcommand{\indexAllW}{23}
\newcommand{\bugzillaIdAllW}{98198}
\newcommand{\numberOfMembersAllW}{15}
\newcommand{\recommendationsAresAllW}{356}
\newcommand{\foundPatternsAresAllW}{15}
\newcommand{\precisionAresAllW}{4}
\newcommand{\recallAresAllW}{100}
\newcommand{\accuracyTokenAresMinAllW}{76}
\newcommand{\accuracyCharactersAresMinAllW}{79}
\newcommand{\accuracyTokenAresMaxAllW}{95}
\newcommand{\accuracyCharactersAresMaxAllW}{95}
\newcommand{\recommendationsLaseAllW}{0}
\newcommand{\foundPatternsLaseAllW}{0}
\newcommand{\precisionLaseAllW}{0}
\newcommand{\recallLaseAllW}{0}
\newcommand{\accuracyTokenLaseAllW}{-100}
\newcommand{\accuracyCharactersLaseAllW}{-100}
\newcommand{\numberOfMembersTwoFilesAvg}{6}
\newcommand{\recommendationsAresTwoFilesAvg}{7}
\newcommand{\foundPatternsAresTwoFilesAvg}{5}
\newcommand{\precisionAresTwoFilesAvg}{90}
\newcommand{\recallAresTwoFilesAvg}{76}
\newcommand{\accuracyTokenAresMinTwoFilesAvg}{92}
\newcommand{\accuracyCharactersAresMinTwoFilesAvg}{92}
\newcommand{\accuracyTokenAresMaxTwoFilesAvg}{96}
\newcommand{\accuracyCharactersAresMaxTwoFilesAvg}{96}
\newcommand{\recommendationsLaseTwoFilesAvg}{9}
\newcommand{\foundPatternsLaseTwoFilesAvg}{5}
\newcommand{\precisionLaseTwoFilesAvg}{92}
\newcommand{\recallLaseTwoFilesAvg}{87}
\newcommand{\accuracyTokenLaseTwoFilesAvg}{78}
\newcommand{\accuracyCharactersLaseTwoFilesAvg}{65}
\newcommand{\recommendationsAresAllAvg}{24}
\newcommand{\foundPatternsAresAllAvg}{6}
\newcommand{\precisionAresAllAvg}{82}
\newcommand{\recallAresAllAvg}{100}
\newcommand{\accuracyTokenAresMinAllAvg}{87}
\newcommand{\accuracyCharactersAresMinAllAvg}{88}
\newcommand{\accuracyTokenAresMaxAllAvg}{94}
\newcommand{\accuracyCharactersAresMaxAllAvg}{94}

\newcommand{\TimeARESTWOFILESCREATEMedian}{0.054162}
\newcommand{\TimeARESTWOFILESCREATEUpperQuartile}{0.681364}
\newcommand{\TimeARESTWOFILESCREATELowerQuartile}{0.016469}
\newcommand{\TimeARESTWOFILESCREATEUpperWhisker}{1.678706}
\newcommand{\TimeARESTWOFILESCREATELowerWhisker}{0.000000}
\newcommand{\TimeARESTWOFILESCREATEMax}{4.163283}
\newcommand{\TimeARESALLCREATEMedian}{2.694334}
\newcommand{\TimeARESALLCREATEUpperQuartile}{13.417884}
\newcommand{\TimeARESALLCREATELowerQuartile}{0.199894}
\newcommand{\TimeARESALLCREATEUpperWhisker}{33.244870}
\newcommand{\TimeARESALLCREATELowerWhisker}{0.000000}
\newcommand{\TimeARESALLCREATEMax}{90.654628}
\newcommand{\TimeARESTWOFILESSEARCHMedian}{19.163743}
\newcommand{\TimeARESTWOFILESSEARCHUpperQuartile}{32.029636}
\newcommand{\TimeARESTWOFILESSEARCHLowerQuartile}{13.255603}
\newcommand{\TimeARESTWOFILESSEARCHUpperWhisker}{60.190687}
\newcommand{\TimeARESTWOFILESSEARCHLowerWhisker}{0.000000}
\newcommand{\TimeARESTWOFILESSEARCHMax}{0.000000}
\newcommand{\TimeARESALLSEARCHMedian}{19.908458}
\newcommand{\TimeARESALLSEARCHUpperQuartile}{32.685366}
\newcommand{\TimeARESALLSEARCHLowerQuartile}{12.806922}
\newcommand{\TimeARESALLSEARCHUpperWhisker}{62.503032}
\newcommand{\TimeARESALLSEARCHLowerWhisker}{0.000000}
\newcommand{\TimeARESALLSEARCHMax}{0.000000}
\newcommand{\TimeARESLASESEARCHMedian}{12.680506}
\newcommand{\TimeARESLASESEARCHUpperQuartile}{19.799619}
\newcommand{\TimeARESLASESEARCHLowerQuartile}{8.997748}
\newcommand{\TimeARESLASESEARCHUpperWhisker}{36.002425}
\newcommand{\TimeARESLASESEARCHLowerWhisker}{0.000000}
\newcommand{\TimeARESLASESEARCHMax}{56.441301}
\newcommand{\TimeLASECREATEMedian}{2.158668}
\newcommand{\TimeLASECREATEUpperQuartile}{3.113988}
\newcommand{\TimeLASECREATELowerQuartile}{1.698019}
\newcommand{\TimeLASECREATEUpperWhisker}{5.237942}
\newcommand{\TimeLASECREATELowerWhisker}{0.000000}
\newcommand{\TimeLASECREATEMax}{10.124570}
\newcommand{\TimeLASESEARCHMedian}{5.150711}
\newcommand{\TimeLASESEARCHUpperQuartile}{7.769817}
\newcommand{\TimeLASESEARCHLowerQuartile}{4.577096}
\newcommand{\TimeLASESEARCHUpperWhisker}{12.558899}
\newcommand{\TimeLASESEARCHLowerWhisker}{0.000000}
\newcommand{\TimeLASESEARCHMax}{13.907895}
\newcommand{\percentDetermineOrderAllCreate}{62}
\newcommand{\percentTreeDifferencingAllCreate}{25}
\newcommand{\percentDetermineOrderTwoFileCreate}{0}
\newcommand{\percentTreeDifferencingTwoFileCreate}{55}
\newcommand{\percentParsingAllSearch}{26}
\newcommand{\percentTreeDifferencingAllSearch}{2}
\newcommand{\percentCreateRecommendationAllSearch}{0}
\newcommand{\percentParsingTwoFilesSearch}{26}
\newcommand{\percentTreeDifferencingTwoFilesSearch}{3}
\newcommand{\percentCreateRecommendationTwoFilesSearch}{0}
\newcommand{\percentParsingTwoFilesLaseSearch}{8}
\newcommand{\percentTreeDifferencingTwoFilesLaseSearch}{3}
\newcommand{\percentCreateRecommendationTwoFilesLaseSearch}{0}

\begin{abstract}
During the life span of large software projects,
developers often apply the same
code changes to different code locations in slight variations.
Since the application of these changes to all locations 
is time-consuming
and error-prone, tools exist that learn
change patterns from input examples,
search for possible pattern applications, and generate 
corresponding recommendations. In many cases, the generated recommendations
are syntactically or semantically wrong due to
code movements in the input examples. Thus, they are of low accuracy
and developers cannot directly copy them into their 
projects without adjustments.

We present the Accurate REcommendation System (ARES) that achieves
a higher accuracy than other tools because its algorithms
take care of code movements when
creating patterns and recommendations.
On average, the recommendations by ARES have an accuracy of 96\% 
with respect to code changes that developers have manually performed in commits of source code archives.
At the same time ARES achieves precision and recall values that are 
on par with other tools.
\end{abstract}

%
%

\ccsdesc[500]{Information systems~Recommender systems}
\ccsdesc[500]{Software and its engineering~Software maintenance tools}

\keywords{Program transformation, refactoring, recommendation system}

\maketitle
\section{Introduction}
Developers often perform the error-prone and repetitive task
of applying the same \textit{systematic edits} to many locations
in their code base. The reasons for
such systematic changes vary. They include bug fixes,
the adaption of call sites to new APIs, etc., and
constitute the majority of structural edits
in software projects~\cite{KIM2009_LSDIFF,NGUYEN2010}.

Developers usually perform systematic edits manually because
often the code is slightly different in each location (with respect to names, contexts, etc.) 
so that they cannot use a plain textual search-and-replace.
Refactoring wizards of IDEs also just offer a set of predefined transformations and thus
only provide help for a subset of all systematic changes. Furthermore,
they do not support code changes that lead to different semantics, which is
necessary in many situations (e.g., to fix errors).

\begin{figure}[b]
\vspace{-0.3cm}
    \captionsetup[subfigure]{justification=centering}
    \subcaptionbox{Example change.}
    {
        \begin{minipage}{0.4\linewidth}
\begin{lstlisting}[basicstyle=\ttfamily\bfseries\scriptsize,language=diff]
  init();

- foo.someMethod(42);
- print(foo);
+ if (foo != null) {
+   foo.someMethod(42);
+   print(foo);
+ }
\end{lstlisting}
        \end{minipage}
        \vspace{-0.2cm}
    }
    \subcaptionbox{Example change.}
    {
        \begin{minipage}{0.4\linewidth}
            \vspace{0.1cm}
\begin{lstlisting}[basicstyle=\ttfamily\bfseries\scriptsize,language=diff]
  init();
- assert (foo != null);
- foo.someMethod(42);
- foo.print();
+ if (foo != null) {
+   foo.someMethod(42);
+   foo.print();
+ }
\end{lstlisting}
        \end{minipage}
        \vspace*{-0.2cm}
    }
    \subcaptionbox{Matching code location.}
    {
    \hspace{2cm}
        \begin{minipage}{0.5\linewidth}
            \vspace{0.1cm}
            \begin{center}
            \hspace{2cm}
\begin{lstlisting}[basicstyle=\ttfamily\bfseries\scriptsize,language=diff]
init();
assert (foo != null);
foo.someMethod(42);
foo.run();
\end{lstlisting}
            \end{center}
        \end{minipage}
        \vspace*{-0.2cm}
    }
    \subcaptionbox{Inaccurate \\recommendation by LASE.}
    {
    \begin{minipage}{0.45\linewidth}
            \vspace{0.1cm}
\begin{lstlisting}[basicstyle=\ttfamily\bfseries\scriptsize,language=java2]
init();
?assert(foo != null);
if (foo != null) {
  foo.someMethod(42);
}
?foo.run();
\end{lstlisting}
        \end{minipage}
        \vspace*{-0.2cm}
    }
    \subcaptionbox{More accurate \\recommendation by ARES.}
    {
    \begin{minipage}{0.45\linewidth}
\begin{lstlisting}[basicstyle=\ttfamily\bfseries\scriptsize,language=java2]
init();
if (foo != null) {
  foo.someMethod(42);
  ?foo.run();
}
\end{lstlisting}
        \end{minipage}
        \vspace*{-0.2cm}
    }
    \vspace*{-0.2cm}
    \caption{Example for code recommendations.}
    \label{fig:example_changes}
    \vspace*{-0.10cm}
\end{figure}

Current research tools~\cite{REFAZER, MENG2013} learn systematic edits from one or more manually provided training examples
and build generalized patterns from them. Then the tools search
the code base for
locations to apply these patterns to and present the code of the applied patterns as recommendations to developers.
Since the tools construct the suggested code (to a large extent) purely syntactically, the code is often wrong
(uses undefined variables, calls non-existing functions, misses statements, etc.).
Thus, the recommended code is often inaccurate, i.e.,
it is not what developers would have written if they had done the systematic edits by hand.
Although such inaccurate recommendations are helpful as they identify locations for code changes,
developers still need to manually adjust them before they can insert them into their projects.

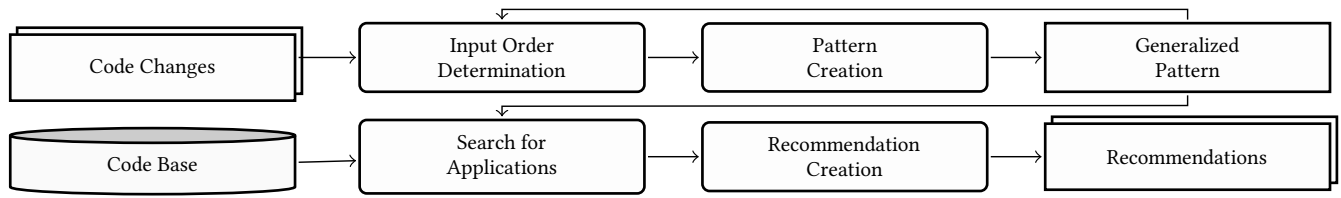
\begin{figure*}[t]
    \begin{minipage}{\linewidth}
    {
        \begin{center}
            \resizebox{1.0\textwidth}{!}
            {
                \begin{tikzpicture}
                    \tikzstyle{component}=[font=\scriptsize, outer sep=1pt, inner sep=1pt, minimum width=2.8cm]
                    \tikzstyle{database}=[component, cylinder, aspect=0.15, shape border rotate=90,
                                          draw=black, line width=0.75pt,
                                          cylinder uses custom fill,
                                          cylinder end fill=black!20, cylinder body fill=black!2,
                                          minimum height=0.65cm]
                    \tikzstyle{action}=[component, rounded corners=2pt,
                                        draw=black, line width=0.75pt, fill=black!1,
                                        inner xsep=8pt, inner ysep=4pt,
                                        text centered, minimum height=0.45cm]
                    \tikzstyle{file}=[component,
                                        draw=black, line width=0.75pt, fill=black!1,
                                        inner xsep=8pt, inner ysep=4pt,
                                        text centered, minimum height=0.45cm]
		    \tikzstyle{files}=[component, copy shadow,
                                        draw=black, line width=0.75pt, fill=black!1,
                                        inner xsep=8pt, inner ysep=4pt,
                                        text centered, minimum height=0.65cm]
                    \tikzstyle{result}=[component, rounded corners=2pt,
                                        draw=black, line width=0.75pt, fill=black!1,
                                        inner xsep=8pt, inner ysep=7pt,
                                        text centered, minimum height=0.45cm, dashed]
                    \tikzstyle{tool}=[component, rectangle, draw=black, line width=0.75pt, fill=black!5,
                                      inner xsep=6pt, inner ysep=12pt,
                                      text centered]
                    \tikzstyle{arrow caption}=[font=\footnotesize, outer sep=-1pt]
                    
                    \node[files] (examples) {\strut Code Changes};
                    \node[action, align=center] (determine input pair) at ([shift={(2.0cm,0.10cm)}]examples.east) {Input Order \\Determination};
                    \node[action, align=center, right=0.5cm of determine input pair] (createpattern) {Pattern \\Creation};
                    \node[file, align=center, right=0.5cm of createpattern] (genpattern) {Generalized \\Pattern};
                    \node[action, align=center, below=0.58cm of determine input pair] (search applications) at ([shift={(2.0cm,0.10cm)}]examples.east) {Search for \\Applications};
                    \node[action, align=center, right=0.5cm of search applications] (createRecommendations) {Recommendation \\Creation};
                    \node[files, align=center, right=0.5cm of createRecommendations] (recommendations) {Recommendations};

                    \node[database, align=center, below=0.20cm of examples] (projects) {\strut Code Base};

                    \draw[->] ($(examples.east) + (0,0.10cm)$) -- node[arrow caption, anchor=south] {} node[arrow caption, anchor=north] (cap_changes) {} (determine input pair);
                    \draw[->] (determine input pair) -- node[arrow caption,anchor=south] {} node[arrow caption,anchor=north] (cap_matrix) {} (createpattern);
                    \draw[->] (createpattern) -- node[arrow caption, anchor=south] {} node[arrow caption, anchor=north] (cap_groups) {} ($(genpattern.west) + (0,0.0cm)$);
                    \draw[->] (genpattern.north) -- + (0,0.10cm) -- ($(determine input pair.north) + (0,0.10cm)$) -- ($(determine input pair.north)$);
		    \draw[->] (genpattern.south) -- + (0,-0.10cm) -- ($(search applications.north) + (0,0.10cm)$) -- ($(search applications.north)$);
                    \draw[->] ($(projects.east) + (0,0.04cm)$) -- node[arrow caption, anchor=south] {} node[arrow caption, anchor=north] (cap_changes) {} (search applications);
                    \draw[->] (search applications) -- node[arrow caption,anchor=south] {} node[arrow caption,anchor=north] (cap_matrix) {} (createRecommendations);
                    \draw[->] (createRecommendations) -- node[arrow caption, anchor=south] {} node[arrow caption, anchor=north] (cap_groups) {} ($(recommendations.west) + (0,0.0cm)$);
                \end{tikzpicture}
            }
        \end{center}
        \vspace*{-0.35cm}
    }
    \end{minipage}
    \caption{Workflow of ARES for one input set of code changes.}
    \label{fig:workflow}
    \vspace*{-0.35cm}
\end{figure*}

Fig.~\ref{fig:example_changes} illustrates this problem.
The two recommendation systems LASE~\cite{MENG2013} and ARES
learn a pattern from the code changes
given in Fig.~\ref{fig:example_changes}(a+b) (in a classic \textit{diff} representation).
Both tools find the matching code location in 
Fig.~\ref{fig:example_changes}(c). LASE suggests the code change
in Fig.~\ref{fig:example_changes}(d). ARES produces the more accurate one in Fig.~\ref{fig:example_changes}(e)
that is closer to what a developer would have written.
There are two reasons for the differences.
First, since change (b) explicitly removes the \textit{assert} statement,
it should not be part of the recommendation.
However, as LASE only applies the common subset of the code
transformations that are present in the training examples,
the \textit{assert} remains untouched. Second,
LASE leaves \textit{foo.run()} in place.
This is wrong because, from a developer's point of view,
\textit{someMethod} and the code in the following line are both \textit{moved}
into the new \textit{if}-statement.
However, many approaches (like LASE) either do not
express code movements accurately or
cannot handle them due to the type of patterns and algorithms they use~\cite{REFAZER, SANTOS2017}.
Since the latter only support
\textit{delete}, \textit{insert}, and \textit{update} operations on the code,
they can only learn to insert \textit{print(foo)} and/or \textit{foo.print()},
but not the more accurate code movement.

To avoid these two sources of inaccuracy, the pattern representation of ARES can both express
variations in the input code changes and code movements. ARES also uses algorithms that can generate
more accurate recommendations based on these patterns.

Fig.~\ref{fig:workflow} shows the workflow of ARES.
The loop in the upper row derives and refines a generalized
pattern that represents all the code changes in a training set.
The loop starts with two code changes from the input set and generates a
pattern for them. Subsequent iterations refine this pattern by considering the
next examples successively.
With a \textit{Generalized Pattern} ARES 
browses a given code base for locations where the pattern is applicable
(\textit{Search for Applications}).
It then applies the transformation encoded in the pattern to a
copy of each found location (\textit{Recommendation Creation})
and presents the transformed copy as recommendation to the user.

Sec.~\ref{sec:design} explains the pattern design and how this helps
in creating more accurate recommendations. Secs.~\ref{sec:order}-\ref{sec:apply} describe the
pattern creation, the search for locations to apply them, and the generation of the
recommendations. We then quantitatively compare ARES with LASE in Sec.~\ref{sec:evaluation}, and
discuss related work before we conclude.

\section{Pattern Design}
\label{sec:design}
ARES uses a pattern representation that is close to source code. Fig.~\ref{fig:pattern} holds an
example pattern (based on our previous work on this topic~\cite{DVP12}) that expresses (and
generalizes) code changes applied to a code base. The set of code changes is the input of ARES.
There are two plain Java code blocks in the pattern, one original code block that represents the
input examples before their transformation and one modified code block that represents the
examples after their transformation. To express variations in the input examples, the patterns
use a set of annotations, added as Java comments.

The example pattern
starts with the \textit{match} annotation that simply declares
the beginning of the pattern. The tags \textit{original} and
\textit{modified} allow an easy distinction between
the two code parts of the pattern for a human reader.
The original part of the \textit{match}
annotation also contains the letter \textit{k}. This \textit{k}
stands for an identifier that occurs in the pattern body and means that
the identifier name of \textit{k} is not fixed and any variable name in a
code location is an acceptable replacement for \textit{k}. In contrast, 
any location to which the pattern is applicable has to use the identifier
name \textit{foo} (line 12). The list of identifiers in the \textit{match}
annotation is one mechanism of ARES to express a generalization.
When ARES creates a recommendation it replaces \textit{k} with the
actual variable name at the respective code location. This increases the
accuracy of the generated recommendations.

\begin{figure}[b]
\vspace{-0.3cm}
		\begin{subfigure}[t]{.23\textwidth}
\begin{lstlisting}[numbers=left, language=java,numbersep=2pt,basicstyle=\ttfamily\scriptsize]
//#match (original, (k)) {
  //#wildcard expr(A1,verbose,1)
  this.init(verbose);
  this.shutdown();
  updateValue();
  k = 0;
  while (k < 10) {
    //#wildcard stmt(A2);
    k++;
  }
  //#wildcard stmt(A3);
  foo.someMethod(42);





//# }





\end{lstlisting}
	\vspace*{-0.35cm}
      \caption{Original part.}\label{fig:motivationFind}
		\end{subfigure}
		\begin{subfigure}[t]{.2\textwidth}
\begin{lstlisting}[language=java,numbersep=2pt,basicstyle=\ttfamily\scriptsize]
//#match (modified) {
  //#use (A1,verbose,1)
  this.init(verbose);
  updateValue();
  for (k = 0; k < 10; k++) {
    //#use (A2);
  }
  if (foo != null) {
    foo.someMethod(42);
  }  
  //#choice {
    //#case
      System.out.print(foo);
    //#case
      this.print(foo);
  //# }
  this.shutdown();
//# } 
\end{lstlisting}
	\vspace*{-0.35cm}
      \caption{Modified part.}\label{fig:motivationLess}
		\end{subfigure}
        \vspace*{-0.35cm}

		\caption{Generalized pattern.}
		\label{fig:pattern}
\end{figure}

Another generalization mechanism is the \textit{wildcard} annotation.
It matches
arbitrary code during the search for suitable code locations.
There are 
two different versions. First, wildcards tagged with \textit{stmt} 
accept none or arbitrary
statements at the code location (see lines 8 and 11). This design provides a solution to
the problem of the deleted \textit{assert} in the introductory example
(Fig.~\ref{fig:example_changes}) as it accepts the \textit{assert} if
it is present and has no effect otherwise. Thus,
\textit{stmt}-wildcards handle variations in the training examples
and increase the accuracy of the generated recommendations.

Second, wildcards tagged with \textit{expr} always refer to the following statement.
They specify which part of
it can contain an arbitrary expression.
It is possible to have several such wildcards referring to the same statement.
For example, the \textit{expr}-wildcard in line 2 specifies that at the first
occurrence of \textit{verbose}
in line 3 the search algorithm of ARES can allow arbitrary
expressions. This means that
at a possible code location for the pattern, the call to \textit{init} may
have none or an arbitrary number of arguments.
This handles variations in the training examples on an even finer
level.

The \textit{modified} part of the pattern does not contain \textit{wildcard} but
\textit{use} annotations.
During the creation of a recommendation ARES replaces the
\textit{use} annotation with the code that was matched by the corresponding wildcard.
A \textit{wildcard} and a \textit{use} correspond to each other if
they have the same name (e.g., \textit{A1}).
As the name can appear anywhere in the modified part,
a pattern can express movements of arbitrary code. This solves
the accuracy problem of the moved \textit{print} methods in the introductory example.

The \textit{modified} part of the pattern also contains a \textit{choice} annotation. ARES creates
this annotation if some training examples add different code. As the added statements have no
connection to the original code, ARES (and also other tools) cannot decide which statements lead
to the most accurate recommendation. ARES instead lets the developer choose among the variants
of the same recommendation. Thus, the \textit{choice} annotation allows ARES to handle additional
variations in the training examples to increase accuracy.

The above discussion demonstrates that our pattern design can increase the accuracy
of recommendations. How to 
generate such patterns from examples and how to use them to create accurate
recommendation is covered in the next sections.

\begin{figure*}[t]
		\begin{subfigure}[t]{.24\textwidth}
\begin{lstlisting}[numbers=left, language=java,numbersep=2pt,basicstyle=\ttfamily\scriptsize]
{
  d = 1.0;
  this.init(true);
  this.shutdown();
  updateValue();
  j = 0;
  while (j < 10) {
    String tmp = "bar";
    this.init(verbose, tmp);
    j++;
  }
  assert (foo != null); 
  
  foo.someMethod(23);





}
\end{lstlisting}
      \caption{Original method \textit{o1} of \textit{c1}.}\label{fig:originalone}
		\end{subfigure}
		\begin{subfigure}[t]{.24\textwidth}
\begin{lstlisting}[language=java,numbersep=2pt,basicstyle=\ttfamily\scriptsize]
{
  d = 1.0;
  this.init(true);
  
  updateValue();

  for (j = 0; j < 10; j++) {
    String tmp = "bar";
    this.init(verbose, tmp);

  }

  if (foo != null) {
     foo.someMethod(23);
  }
  this.print(foo);
  this.shutdown();


}
\end{lstlisting}
      \caption{Modified method \textit{m1} of \textit{c1}.}\label{fig:modifiedone}
		\end{subfigure}
		\begin{subfigure}[t]{.24\textwidth}
\begin{lstlisting}[language=java,numbers=left, numbersep=2pt,basicstyle=\ttfamily\scriptsize]
try {
  this.i = 5;
  this.init(verbose);
  this.shutdown();
  updateValue();
  k = 0;
  while (k < 10) {
    updateValue();
    printValue("foo");
    k++;
  }


  foo.someMethod(42);



} catch (Exception e) {

}
\end{lstlisting}
      \caption{Original method \textit{o2} of \textit{c2}.}\label{fig:originaltwo}
		\end{subfigure}
	\begin{subfigure}[t]{.24\textwidth}
\begin{lstlisting}[language=java,numbersep=2pt,basicstyle=\ttfamily\scriptsize]
try {
  this.i = 5;
  this.init(verbose);
  
  updateValue();
 
  for (k = 0; k < 10; k++) {
    updateValue();
    printValue("foo");

  }

  if (foo != null)  {
     foo.someMethod(42);
  }
  System.out.print(foo);
  this.shutdown();
} catch (Exception e) {
  this.shutdown();
}
\end{lstlisting}
      \caption{Modified method \textit{m2} of \textit{c2}.}\label{fig:modifiedtwo}
		\end{subfigure}
		\vspace*{-0.3cm}

		\caption{Code changes \textit{c1} and \textit{c2}.}
		\label{fig:example}
    \vspace*{-2.0mm}
\end{figure*}

\section{Input Order Determination}
\label{sec:order}

When constructing the generalized pattern, ARES considers the input code
changes one after the other in the loop that is shown in the upper half of
Fig.~\ref{fig:workflow}. The order in which ARES uses them influences the
pattern. Two changes that are very different
(i.e., that have a large edit distance) probably lead to a generalized pattern
that makes excessive use of wildcards to
hide away the differences. The resulting pattern is over-generalized and will
match in many locations of the
code base. The smaller the edit distance between two changes is, the smaller
are the differing code fragments
that ARES hides in wildcards. Hence, the key idea is that in every iteration of
the loop, ARES first identifies
a code change that is as close to the current working generalized pattern as
possible. Initially, in the first iteration when there is no working generalized pattern 
ARES chooses the two code changes that are as close to 
all other changes in the input set as possible.

To determine the edit distance, ARES uses a
tree differencing algorithm that extracts the differences between two abstract
syntax trees (ASTs). In general, this is more precise than approaches based on
strings or tokens. Thus, ARES uses a tree differencing algorithm to
extract code differences throughout the whole process (unless stated
otherwise). Tree differencing algorithms differ with respect to their precision in tracking
code movements. A tree differencing algorithm that reliably detects code
movements leads to an edit distance that better captures the closeness of two
input examples, which is important for the input order determination.
Thus, ARES uses MTDIFF~\cite{MTDIFF}, the currently most precise 
tree differencing algorithm that considers code movements.

Let us sketch this input ordering process by means of an example.
Assume that there are four
code changes $c1$..$c4$. Each of them consists of
an \underline{o}riginal method block $oi$ and a \underline{m}odified method block $mi$.
To obtain the distance between two code changes $ci$ and $cj$, ARES computes the number
of edit operations required to transform $oi$ into $oj$ plus the number of edit
operations to transform $mi$ into $mj$. 
Table~\ref{fig:eop_matrix}
holds some fictitious edit distances for this example. 
As initial pair, ARES selects the examples that represent
the two columns with the lowest sum ($c2$ and $c3$ in the example). The first iteration
then constructs a working \textit{Generalized Pattern}.
Sec.~\ref{sec:patterncreation} below describes this step in detail. The next
iteration of the loop then computes the edit distances from the working generalized pattern
to the remaining code changes ($c1$, $c4$)
and uses the change with the smallest distance to it for the next iteration.

Picking input examples in an order that yields fewer wildcards increases the
accuracy of the recommendations
as wildcards can hide code transformations.
For example, it is possible that the combination of the working pattern with
an input example forces ARES to generalize the loops in Fig.~\ref{fig:pattern} into an
extra wildcard.
The resulting pattern would still match relevant code locations
in the search step, but ARES would no longer transform the \textit{while}
into a \textit{for} loop. Thus, more wildcards lead to less accurate recommendations.

\begin{table}[t]
		\centering
		\scriptsize
		\caption{Edit distances between code changes \textit{c1}..\textit{c4}.}
		\label{fig:eop_matrix}
		\setlength{\tabcolsep}{5pt}
		\begin{tabular}{c|cccc}
						        & $c1$        & $c2$      & $c3$      & $c4$
				\\\hline
			\parbox[c][2.2em]{1.4em}{$c1$}   & -          & $4$        & $5$          & $6$        \\
			\parbox[c][2.2em]{1.4em}{$c2$}   & $4$        & -          & $2$          & $2$        \\
			\parbox[c][2.2em]{1.4em}{$c3$}   & $5$        & $2$        & -            & $2$        \\
			\parbox[c][2.2em]{1.4em}{$c4$}   & $6$        & $2$        & $2$          & -    \\\hline
			\parbox[c][2.2em]{1.4em}{$\sum$}  & $15$       & $8$        & $9$          & $10$ \\

		\end{tabular}
\vspace{-0.3cm}
\end{table}

\section{Pattern Creation}
\label{sec:patterncreation}

The input of the \textit{Pattern Creation} step are two code changes ($c1$ and $c2$)
with their respective original and modified method bodies.
Fig.~\ref{fig:example} holds the running example for this 
section, from which ARES creates the pattern in Fig.~\ref{fig:pattern}.
As discussed in Sec.~\ref{sec:design},
this pattern solves the accuracy problems.

The \textit{Pattern Creation} step adds \textit{wildcard}
annotations where the original method bodies $o1$ and $o2$ differ. Similarly, it adds
\textit{use} annotations where $m1$ and $m2$ differ. To do so, ARES uses the tree differencing
algorithm MTDIFF which takes one original and one modified AST as input and then matches nodes
from the original AST with nodes of the modified AST. Two nodes are a possible match if they have
the same type, e.g., if both are identifiers. MTDIFF uses heuristics for this matching of nodes.
Based on the node matches, MTDIFF generates a small \textit{edit script}, i.e., a list of edit
operations that transform the original AST into the modified AST. It uses four different
types of edit operations on the granularity of AST nodes, namely \textit{delete}, \textit{insert},
\textit{update}, and \textit{move}. The \textit{move} operation moves a complete subtree to a new
location. 

Any AST node that is part of such an edit operation identifies a change between the method bodies
that the \textit{Pattern Creation} handles with annotations. It
also keeps all the remaining code unchanged in the pattern to increase the accuracy.

Below we show in detail how ARES uses the edit scripts \textit{D(o1,o2)} and \textit{D(m1,m2)}
provided by MT\-DIFF to create the pattern. To determine the correct names in the \textit{use}
annotations ARES also requires \textit{D(o1,m1)} and \textit{D(o2,m2)}. As this process is the
most complex part of ARES, we present it in six steps.

\subsection{Change Isolation}
When ARES generates a pattern from the input examples, it has to make sure that the edit scripts
only cover the sections of the code that contain the actual change. Surrounding code that has
nothing to do with the change but still is different in the input examples should not be part of
the pattern. If there was no change isolation, such surrounding code would cause an
over-generalized pattern and thus inaccurate recommendations. In Fig.~\ref{fig:example} change
$c1$ does not have a \textit{try} statement whereas $c2$ does. Without a preceding isolation of
the main change, MTDIFF would create an \textit{insert} operation for the \textit{try} statement
that in turn results in a \textit{wildcard} annotation. This would lead to an
over-generalized pattern with a single wildcard. To avoid this, ARES isolates the code parts that actually
contain the relevant change, e.g., the body of the \textit{try} node.

ARES implements the necessary \textit{Change Isolation} in two phases. The first phase identifies
the lowest nodes in the ASTs that encapsulate the changes, the second phase applies several
heuristics if the first phase still leads to an over-generalization. In the first phase, ARES
works with the ASTs of the original method $o_i$ and the corresponding modified method $m_i$ of a
change. The edit script \textit{D(o$_i$, m$_i$)} helps to identify the lowest root nodes in the two ASTs
that are affected by all the edit operations. Each such change-root has the maximal distance from the
root node of its AST and still encapsulates all the differences between $o$ and $m$.
In the example, the change-root of $o1$/$m1$ is the
complete method block from line 1 to 20. The change-root of $o2$/$m2$ is the \textit{try} node (due to
the change in the \textit{catch}). As the roots are of different types, selecting these two
nodes would still lead to an over-generalization. This and similar input examples are addressed by
phase two. In all easier cases, ARES uses the change-roots for isolation
and pattern generation to exclude
all surrounding code.

If the first phase cannot prevent an over-generalization, the second phase applies the following
three heuristics in the described order: (a) Search for similar statements (i.e., statements
paired together by MTDIFF) in the children of the change-root of $o1$ and the children of the change-root of $o2$.
For $m1$ resp. $m2$, there have to be matching nodes in \textit{D(o1,m1)} resp. \textit{D(o2,m2)}.
(b) Reverse the roles of $c1$ and $c2$ and then try (a) again. (c) Performs (a)
and (b) again but this time with the grandchildren instead of the children. If all three
heuristics fail, the current implementation of ARES stops without a generated pattern to reduce
the execution time.

On the example, phase two identifies similar statements in the block of $c1$ and in the body of
the \textit{try} statement. These nodes isolate the actual change, avoid over-generalization in the
resulting pattern, and thus reduce the irrelevant and inaccurate recommendations.

\subsection{Edit Script Adjustment}
While \textit{Change Isolation} avoids over-generalization, \textit{Edit Script Adjustment}
finds a balance between over-generalization and over-fitting (i.e., the
creation of patterns that only match in very specific situations, e.g., the training set). To
keep a balance, ARES uses a rule-based system with over 50 rules (rule description
 on Github: \url{https://github.com/FAU-Inf2/ARES})---far too many to present individually within the
space restrictions of this paper. Hence, we can only discuss the two main issues
that influence the balance and apply the relevant rules to the running example.

The first issue is that in most cases the tree differencing is too fine-grained. It often
identifies many small changes on the level of expressions but only few changes on the level of
statements, especially if the training examples are quite similar. The resulting patterns
then have \textit{wildcards} and \textit{uses} for fine-grained expressions that often only fit
the training examples. The accuracy of the recommendations will be high, but the recall will be
low.

To avoid this, ARES looks for matching statements that vary in many of their sub-expressions. In
those situations, ARES adjusts the MTDIFF-generated edit script to use a single edit operation
that covers the whole statement (instead of many edit operations for all the differing
sub-expressions). For example, rule \#48 (see Fig.~\ref{fig:ruleexample}) adjusts the edit script
for a declaration statement. It applies to line 2 in Fig.~\ref{fig:example}. Although both the
left and the right hand side of the assignment differ in $o1$ and $o2$, MTDIFF keeps the
assignment as this keeps the size of the edit script \textit{D(o1,o2)} small. To avoid
over-fitting, ARES adjusts the edit script and replaces all \textit{delete} and \textit{insert}
operations for both the left and the right hand side of the assignments (lines 8--9 in
Fig.~\ref{fig:ruleexample}) with one \textit{insert} operation for a full statement 
(line 10 in Fig.~\ref{fig:ruleexample}). Adding a \textit{delete} operation for the assignment is
unnecessary as this would only lead to a wildcard at the same position.

\begin{figure}[t]
\hspace{0.5cm}
\footnotesize
 \begin{algorithmic}[1]
\Function{rule48}{editOps, mapping}
\LState declMappings $\leftarrow$ getDeclarationMappings(mapping)
\For {(d$_{o1}$, d$_{o2}$) $\in$ declMappings}
  \LState l$_{o1}$ $\leftarrow$ leftSide(d$_{o1}$); l$_{o2}$ $\leftarrow$ leftSide(d$_{o2}$)
  \LState r$_{o1}$ $\leftarrow$ rightSide(d$_{o1}$); r$_{o2}$ $\leftarrow$ rightSide(d$_{o2}$)
  \If {isDeleted(l$_{o1}$, editOps) $\wedge$ isDeleted(r$_{o1}$, editOps)}
      \If {isInserted(l$_{o2}$, editOps) $\wedge$ isInserted(r$_{o2}$, editOps)}
          \LState removeOpsForT(l$_{o1}$, editOps); removeOpsForT(r$_{o1}$, editOps);
          \LState removeOpsForT(l$_{o2}$, editOps); removeOpsForT(r$_{o2}$, editOps);
	  \LState addInsertForNode(d$_{o2}$, editOps)
      \LEndIf
  \LEndIf
\EndFor \vspace{-0.6ex}
\EndFunction
\end{algorithmic}
\vspace{-0.2cm}
\caption{Edit script adjustment Rule \#48.}
\vspace{-0.3cm}
\label{fig:ruleexample}
\end{figure}

Since finding an optimal list of edit operations that includes
code movements is NP-hard~\cite{BILLE2005,BOOBNA2004}, all move-aware
tree differencing
algorithms rely on heuristics
and there are cases in which the edit script is not optimal.
Any non-optimal list leads to
unwanted wildcards and thus increases the generality unnecessarily.
This is the second balancing issue that the
edit script adjustment addresses.

For instance, Rule \#31
examines moves across nested code blocks. It applies to line 3 in Fig.~\ref{fig:example}.
MTDIFF determines that the number of edit
operations is minimized if the code of line 9 in $o1$ is moved to line 3 in $o2$.
As a consequence, MTDIFF generates a \textit{move} operation. Due to this \textit{move}
operation, ARES would insert a wildcard in line 3.
This is too general. Instead, the adjustment step replaces
the \textit{move} with a \textit{delete} operations for the call in line 9 of $o1$.
Then it adds the \textit{insert} operation for \textit{verbose} in line 3 of $o2$.

Rules \#13--19 handle movements of identical statements.
For example, there exist two identical statements in $o2$
(lines 5 and 8) for the call in line 5 of $o1$.
Although this is not a problem for the
running example, the adjustment step has to take care of
wrong pairs of two identical statements inside the tree differencing results.
Otherwise, there are unnecessary
\textit{moves} in the list of edit operations which can lead to an over-generalized pattern.

Rules \#42--46 handle changes that are already covered by changes of the parent
statement. These rules apply to the lines 8 and 9 of $o1$ that are replaced
by new statements in $o2$.
 MTDIFF generates \textit{delete} and \textit{insert} operations for
each AST node in both lines. For example, MTDIFF generates a \textit{delete}
operation for \textit{String}, \textit{tmp}, etc. For the \textit{Pattern Creation}
only the \textit{insert} for the invocations of \textit{updateValue} and \textit{printValue}
in $o2$ are relevant. Thus, the rules remove
all edit operations on nodes that are part of the statements.
For the \textit{delete} of the \textit{assert} in line 12, ARES also replaces the deleted expressions with
a single \textit{delete} of the complete statement.

\subsection{Match Insertion}
This step inserts the \textit{match} annotation into the generalized pattern (see
Fig.~\ref{fig:pattern}). The main purpose of this annotation is to provide a list of identifier
names that differ between $o1$ and $o2$. Using wildcards for them is too verbose and would
unnecessarily enlarge the patterns. Instead, ARES uses the matched node pairs from
\textit{D(o1,o2)} to find identifiers in the same match pair but with different names. For the
running example, this is only the pair $(j, k)$ of the loop variables. By mentioning the names in
the annotation, ARES can remove them from all \textit{update} operations in both \textit{D(o1,o2)}
and \textit{D(m1,m2)}.

\subsection{Wildcard and Use Insertion}
This step adds the \textit{wildcard} and \textit{use} annotations to the pattern according to the
edit operations that are left after the previous steps. ARES
replaces each statement that an
\textit{insert} adds in $o2$ with a wildcard. If the \textit{insert} operation
adds an expression, ARES adds a \textit{wildcard} annotation in front of the statement.
Similarly, for each \textit{insert} that affects expressions or statements in $m2$ ARES adds
\textit{use} annotations. For each \textit{delete} operation that removes a node from $o1$ ARES
adds a \textit{wildcard} annotation at the corresponding spot in $o2$ (determined with
\textit{D(o1,o2)} and the heuristics). In the same way, ARES replaces each \textit{delete} in $m1$
with a \textit{use} annotation in $m2$. As \textit{move} operations basically delete a node in
$o1$/$m1$ and add it in $o2$/$m2$, ARES adds wildcards for the \textit{delete} and \textit{insert}
operation expressed by the \textit{move}. Note that at this point there are no longer \textit{update}
operations as the edit script adjustment either removed them or replaced them with \textit{insert}
operations. ARES also memorizes which annotation belongs to which edit operation in order to
facilitate the following name assignment step.

Since in the example the \textit{insert} of \textit{verbose} is left in \textit{D(o1,o2)},
ARES adds a \textit{wildcard} annotation before 
the statement and tags it with \textit{expr}. Additionally,
it specifies which expression corresponds to the wildcard (\textit{verbose}).
As it is possible that an expression occurs several times on the same statement,
the wildcard in line 3 of the final pattern in Fig.~\ref{fig:pattern}(a)
also specifies the number of the occurrence
($1$ for the example). Fig.~\ref{fig:pattern}(b) shows the corresponding \textit{use}.

Another operation is the \textit{insert} of the assignment in line 2 of $c2$. ARES replaces
this assignment with a \textit{wildcard} annotation in $o2$. As it is a replacement
for a complete statement,
the tag \textit{stmt} is added
to the wildcard. Similarly, ARES replaces the assignment in $m2$ with
a \textit{use} annotation.
For the inserted statements in lines 8 and 9 ARES also inserts \textit{wildcard}
and \textit{use} annotations. After the insertion of the annotations in
line 9, ARES immediately combines both adjacent annotations into a single one.

ARES proceeds with the remaining edit operations in this fashion and finally creates
the result in Fig.~\ref{fig:pattern}.

\subsection{Wildcard Name Assignment}
This step assigns the names that link the \textit{wildcard} 
annotations in the original part to the \textit{use} annotations in the modified part.
Since these names can occur in different
spots on both sides of a pattern, they encode code movement. Hence,
they are crucial for the accuracy of the recommendations.

To identify the correct \textit{wildcard}/\textit{use} pairs, this step examines the
statements that were replaced by wildcards (memorized by the previous step).
In the running example, a wildcard replaced
\textit{updateValue} in line 8 of $o2$. Then ARES uses \textit{D(o2,m2)} to find 
a node in $m2$ that is matched to the moved \textit{updateValue} from $o2$. In the example,
the matched node is the call of \textit{updateValue} in $m2$. As
\textit{updateValue} in $m2$ was also replaced by a \textit{use}, ARES links the
\textit{wildcard} and \textit{use} of
the call together and gives them the same name (\textit{A2} in the example).
Similarly, ARES assigns the other names to create Fig.~\ref{fig:pattern}.

The previous steps may create a pattern that starts with a \textit{stmt} wildcard. However, when searching
for applications it is unclear which sequence of statements this wildcard should match.
Therefore, ARES enforces that patterns begin with a specific statement instead of a wildcard. In
the running example, ARES removes the \textit{wildcard} that replaced the assignment in line 2 of
$o2$. If there is a corresponding \textit{use} with the same name and this \textit{use} is also
the first statement in the pattern, ARES also removes it. The same applies to \textit{stmt}
wildcards at the end of the pattern as they match the complete remaining function and thus the
recommendation would be unnecessarily large. The evaluation shows that this has no negative
impact on precision and recall compared to other recommendation systems. If the assigned
\textit{use} of a removed wildcard is not at the top or bottom of the pattern, ARES keeps the
\textit{use} annotation and only removes the assigned name. This is necessary since the
corresponding \textit{wildcard} is no longer present.

\subsection{Choice Insertion}
The final step handles differences between the modified parts of the code changes that do not
correspond to code in the original parts. To increase the accuracy of the recommendations it is
necessary to handle those differences explicitly. After the previous steps, the differences are
visible as they correspond to \textit{use} annotations without assigned names. This step replaces
each such \textit{use} with a \textit{choice} annotation because the input examples provide
insufficient information to determine the right recommendation based on a purely syntactical
approach.

In the running example, the \textit{use} that replaced the \textit{assert}
and \textit{print} call has no assigned name and is thus changed into
a \textit{choice} annotation. Based on $c1$ and $c2$ it is impossible
to determine which \textit{print} invocation should be part of the recommendation. Hence,
ARES generates several recommendations, one for each variant.
It is up to the developer to decide which of the recommendations is the most
appropriate. Similar to the name assignment process,
ARES only has to examine the statement that was replaced by the \textit{use}
annotation to identify the code for the \textit{case} annotations
in Fig.~\ref{fig:pattern}(b).

\section{Search for Applications}
\label{sec:search}

This section explains how ARES finds
code locations where a given pattern is applicable.
For a high accuracy, it is important that the algorithm
identifies the correct matches between the AST nodes in the pattern
(including wildcards) and AST nodes at the code location. 
To make the algorithm as fast as possible, this section no longer
relies on a tree structure but uses a serialized list of AST nodes.
However, this is no limitation to the movement support as the 
\textit{Recommendation Creation} (Sec.~\ref{sec:apply}) handles them.

As example we use the pattern in Fig.~\ref{fig:pattern} and browse the code in
Fig.~\ref{fig:application}(a). ARES first searches for suitable starting points in the code and
then executes the AST-node matching from there in parallel. Such a starting point is an AST node
that has the same type as the first AST node in the original part of the pattern and that is not
part of an annotation. The first node in Fig.~\ref{fig:pattern} is the method call in line 3.
Suitable starting points in Fig.~\ref{fig:application}(a) are the calls in lines 2, 3, 4, 7, and
10. ARES uses each of these calls as a \textit{startNode} when executing the algorithm given in
Fig.~\ref{fig:searchalg}. The other input arguments are the AST of the method that contains
\textit{startNode} and the AST of the original part of the pattern.

In lines 2 and 3 of Fig.~\ref{fig:searchalg}, ARES generates a list of AST nodes for the code
location (starting at \textit{startNode}) and a list of AST nodes for the pattern (starting from
the first code node in the body of the \textit{match} annotation). With the loop in lines 6--29
the algorithm compares both lists of AST nodes to determine whether they are identical (with
respect to the wildcards), in which case the pattern is applicable to the code location that
starts at \textit{startNode}.

\begin{figure}[t]
		\begin{subfigure}[t]{.23\textwidth}
\begin{lstlisting}[numbers=left, language=java,showlines=true,numbersep=2pt,basicstyle=\ttfamily\scriptsize]
Foo foo = Library.getObject();
this.init(getVerbose());
this.shutdown();
updateValue();
c = 0;
while (c < 10) {
  System.out.println(c); 
  c++;
}
foo.someMethod(99);
return foo;
\end{lstlisting}
      \caption{Matching code location.}\label{fig:examplelocation}
		\end{subfigure}
		\begin{subfigure}[t]{.2\textwidth}
\begin{lstlisting}[language=java,numbersep=2pt,basicstyle=\ttfamily\scriptsize]
Foo foo = Library.getObject();
this.init(getVerbose);
updateValue();
for (c = 0; c < 10; c++) {
  System.out.println(c);
}
if (foo != null) {
  foo.someMethod(99);
}  
this.shutdown();
return foo;
\end{lstlisting}
      \caption{Recommendation.}\label{fig:examplerecommendation}
		\end{subfigure}
		\vspace*{-0.3cm}
		\caption{Recommendation Creation.}
		\label{fig:application}
				\vspace*{-0.3cm}
\end{figure}

\begin{figure}[b]
\vspace{-0.3cm}
\hspace{0.5cm}
\footnotesize
 \begin{algorithmic}[1]
\Function{search}{startNode, methodBody, originalPatternPart}
\LState NL$_{cl}$ $\leftarrow$ getNodes(methodBody, startNode)
\LState NL$_{p}$ $\leftarrow$ getPatternNodes(originalPatternPart)
\LState pos$_{cl}$ $\leftarrow$ 0; pos$_{p}$ $\leftarrow$ 0
\LState resets $\leftarrow$ $\emptyset$; matches $\leftarrow$ $\emptyset$; visited $\leftarrow$ $\emptyset$
\While {pos$_{cl}$ $<$ $|$NL$_{cl}|$ $\wedge$ pos$_{p}$ $<$ $|$NL$_{p}|$} 
    \LState n$_{cl}$ $\leftarrow$  getNode(NL$_{cl}$, pos$_{cl}$); n$_{p}$ $\leftarrow$  getNode(NL$_{p}$, pos$_{p}$)
    \LState w $\leftarrow$ null
    \If {isWildcard(n$_{p}$)}
        \LState w $\leftarrow$ n$_{p}$
    \ElsIf {hasAssociatedWildcard(n$_{p}$)}
        \LState w $\leftarrow$ getAssociatedWildcard(n$_{p}$)
    \LEndIf \vspace{-0.6ex}
    \If {w $=$ null}
        \If {isMatch(n$_{cl}$, n$_{p}$, originalPatternPart)}
            \LState matches $\leftarrow$ (n$_{cl}$, n$_{p}$)
            \LState pos$_{cl}$ $\leftarrow$ pos$_{cl}$ + 1; pos$_{p}$ $\leftarrow$ pos$_{p}$ + 1
            \LState \textbf{continue}
        \LEndIf
    \Else
        \If {w $\notin$ visited}
            \LState addLast(resets, (pos$_{cl}$, pos$_{p}$ + 1, matches, visited))
            \LState visited $\leftarrow$ visited $\cup$ w
        \LEndIf
        \If {allowedNode(w, n$_{cl}$)}
            \LState matches $\leftarrow$ (n$_{cl}$, w)
            \If {allowedReset(w, n$_{cl}$)}
                \LState addLast(resets, (pos$_{cl}$ + 1, pos$_{p}$ + 1, matches, visited))    
            \LEndIf \vspace{-0.6ex}
            \LState pos$_{cl}$ $\leftarrow$ pos$_{cl}$ + 1
            \LState \textbf{continue}
        \LEndIf
    \LEndIf
    \If {resets = $\emptyset$}
        \textbf{return} null
    \LEndIf
    \LState (pos$_{cl}$, pos$_{p}$, matches, visited) $\leftarrow$ removeLast(resets)
\LEndWhile \vspace{-0.8ex}
\If {pos$_{p}$ $=$ $|$NL$_{p}|$}
    \LState \textbf{return} matches
\Else
    \LState \textbf{return} null
\LEndIf \vspace{-0.6ex}
\EndFunction
\end{algorithmic}
\vspace{-0.2cm}
\caption{Algorithm to search for suitable code locations.}
\label{fig:searchalg}
\end{figure}

In the running example, both lists start with a \textit{call} node. As the node of the pattern is
not a wildcard, \textit{w} (line 13) is \textit{null} and the algorithm uses \textit{isMatch} to
compare n$_{cl}$ of the code location with n$_p$ of the pattern. The function \textit{isMatch}
returns true if the types of these AST nodes are identical (e.g., if both are \textit{call}
nodes). There are only two exceptions. If both AST nodes are identifiers, \textit{isMatch} is true
if n$_p$ is in the list of identifiers of the \textit{match} annotation or if the identifier names
are equal. The second exception concerns boolean constants as they make a large difference in the
pattern due to their limited value range. In this case, \textit{isMatch} is only true if both are
identical. ARES does not compare the values of other constants. This increases the generalization
of a pattern and makes it applicable to more code locations because different literals do not
prevent a pattern from being applicable.

If the comparison with \textit{isMatch} is successful, the algorithm adds the pair
(n$_{cl}$, n$_p$) to the set of matched nodes (\textit{matches}). The recommendation
algorithm later uses this set to replace the identifiers in the pattern (e.g., \textit{k}) with
identifiers of the code location (e.g., \textit{c}) and also to replace constants in the pattern
(e.g., $42$) with constants in the code location (e.g., $99$).
If the comparison with \textit{isMatch} is unsuccessful, the algorithm
looks for another valid position to backtrack (lines 28--29). If there is a
valid position, ARES resets the variables to the saved values and continues from the last valid position.

Backtracking is necessary due to the use of wildcards. The pattern design (see Sec.~\ref{sec:design})
allows a wildcard to replace none or more statements/expressions. This leads to several valid
end points of a wildcard match and creates the reset nodes. In the example,
the node \textit{verbose} is associated with a wildcard. Thus, \textit{getAssociatedWildcard} 
identifies the wildcard in line 3 of the pattern. As this is the first appearance of this
wildcard, the algorithm creates a valid reset position in line 20. This reset position
covers the case that the wildcard is empty and does not match any code at this location.
The reset point starts at the position after \textit{verbose} (pos$_p$ + 1) and thus continues
without a match. 

In our example, \textit{init} has an argument and thus the wildcard has
a node to match. The function \textit{allowedNode} checks whether the wildcard can replace n$_{cl}$.
A replacement is possible if n$_{cl}$ is within the scope of the wildcard. For the wildcard in
line 2, the scope consists of the arguments of \textit{init}. Here this is the case
and the algorithm adds the appropriate match (line 23). If n$_{cl}$ is also the last node of
one argument of \textit{init}, the node is also a valid end point of the wildcard and
the algorithm adds a new reset point that starts at the end of the argument (lines 24--25). 
Then the search continues at the next node in the list of the code location (lines 26--27).
The algorithm handles the other wildcards (lines 8, 11) in a similar fashion. The difference is only
that this time \textit{allowedNode} accepts all nodes from the current code block and
\textit{allowedReset} accepts only complete statements.

After the loop terminates, the algorithm checks whether the process reached the end
of the template and thus whether the code location matches all nodes of the template. In this case
the search is successful and returns the set of matched nodes (lines 30--33).

\section{Recommendation Creation}
\label{sec:apply}

The general idea of the creation step is to use MTDIFF to create the list of edit operations that change the \textit{original} 
part of the pattern into the \textit{modified} part (see Fig.~\ref{fig:pattern})
and to apply these edit operations to a copy of the code location (identified in the previous step).
Note that ARES does not use the \textit{modified} part of the pattern as the base
for the recommendation. Using the list of edit operations instead has the advantage
that ARES can preserve parts of the copied code, e.g., the identifier \textit{c} and
the number \textit{99} (see Fig.~\ref{fig:application}(a)). This
increases the accuracy of the recommendation.

To be able to apply the list of edit operations,
the copied code has to look like the \textit{original} part of the pattern.
For that purpose, ARES removes all AST nodes from the 
copy that were matched with a wildcard during the search.
Then ARES inserts the wildcard annotations into the copy.
After these changes, the copy looks like the \textit{original} part of the pattern in
Fig.~\ref{fig:pattern}(a) except for the preserved code parts.
This allows ARES to apply the edit operations to create
the \textit{modified} part of the pattern (with respect to the preserved code parts).
Then ARES replaces the \textit{use} annotations with the code that
is paired with the corresponding wildcard,
i.e., the wildcard with the same name.

Finally, ARES works on the \textit{choice} annotations. For a single \textit{choice} ARES could
create one copy of the recommendation per \textit{case}. If there are more \textit{choice}s, then
conceptually there is the potential for exponential growth. Hence, ARES limits the number of
copies. If there are at most $max$ case statements in all \textit{choice}s (2 in the example),
ARES creates $max+1$ copies of the current recommendation code. The $n$-th copy contains the code
in the $n$-th \textit{case} annotation of each \textit{choice}. If a \textit{choice} annotation
has fewer than $n$ cases, we omit this choice completely. This implies that the last copy does not
contain any code from \textit{choice} annotations. After this step, ARES presents the $max+1$
copies as variants of one recommendation to the developer.

\begin{table*}[t]
    \caption{Comparison with LASE on 23 code changes from Eclipse JDT and Eclipse SWT.}
	\setlength{\tabcolsep}{4pt}
	\centering
        \scriptsize
	\begin{tabular}{c|c|r||r|r|r|r|r|r||r|r|r|r|r|r||r|r|r|r|r|r}
\multicolumn{3}{c||}{} & \multicolumn{6}{c||}{\textbf{LASE---Two Input Code Changes}} & \multicolumn{6}{c}{\textbf{ARES---Two Input Code Changes}} & \multicolumn{6}{c}{\textbf{ARES---All Code changes}}\\ 
Id & Bugzilla \textit{Id} & $m$ & $\triangle$ & $\checkmark$ & \textit{A$_T$\%} & \textit{A$_C$\%} & \textit{P\%} & \textit{R\%} &  $\triangle$ & $\checkmark$ & \textit{A$_T$\%} & \textit{A$_C$\%} & \textit{P\%} & \textit{R\%} &  $\triangle$ & $\checkmark$ & \textit{A$_T$\%} & \textit{A$_C$\%} & \textit{P\%} & \textit{R\%} \\ 
\hline
\indexTwoFilesA{} & \bugzillaIdTwoFilesA{} & \numberOfMembersTwoFilesA{} & \recommendationsLaseTwoFilesA{} & \foundPatternsLaseTwoFilesA{} & \accuracyTokenLaseTwoFilesA{} & \accuracyCharactersLaseTwoFilesA{} & \precisionLaseTwoFilesA{} & \recallLaseTwoFilesA{} & \recommendationsAresTwoFilesA{} & \foundPatternsAresTwoFilesA{} & \accuracyTokenAresTwoFilesA{}                                       & \accuracyCharactersAresTwoFilesA{}                                            & \precisionAresTwoFilesA{} 		& \recallAresTwoFilesA{} 		& \recommendationsAresAllA{} & \foundPatternsAresAllA{} & \accuracyTokenAresAllA{}                                  & \accuracyCharactersAresAllA{} & \precisionAresAllA{} & \recallAresAllA{} \\
\indexTwoFilesB{} & \bugzillaIdTwoFilesB{} & \numberOfMembersTwoFilesB{} & \recommendationsLaseTwoFilesB{} & \foundPatternsLaseTwoFilesB{} & \accuracyTokenLaseTwoFilesB{} & \accuracyCharactersLaseTwoFilesB{} & \precisionLaseTwoFilesB{} & \recallLaseTwoFilesB{} & \recommendationsAresTwoFilesB{} & \foundPatternsAresTwoFilesB{} & \accuracyTokenAresMinTwoFilesB{}$/$\accuracyTokenAresMaxTwoFilesB{} & \accuracyCharactersAresMinTwoFilesB{}$/$\accuracyCharactersAresMaxTwoFilesB{} & \textbf{\precisionAresTwoFilesB{}} 	& \recallAresTwoFilesB{} 		& \recommendationsAresAllB{} & \foundPatternsAresAllB{} & \accuracyTokenAresAllB{}                                  & \accuracyCharactersAresAllB{} & \precisionAresAllB{} & \recallAresAllB{} \\
\indexTwoFilesC{} & \bugzillaIdTwoFilesC{} & \numberOfMembersTwoFilesC{} & \recommendationsLaseTwoFilesC{} & \foundPatternsLaseTwoFilesC{} & \accuracyTokenLaseTwoFilesC{} & \accuracyCharactersLaseTwoFilesC{} & \precisionLaseTwoFilesC{} & \recallLaseTwoFilesC{} & \recommendationsAresTwoFilesC{} & \foundPatternsAresTwoFilesC{} & \accuracyTokenAresTwoFilesC{}                                       & \accuracyCharactersAresTwoFilesC{}                                            & \precisionAresTwoFilesC{} 		& \recallAresTwoFilesC{} 		& \recommendationsAresAllC{} & \foundPatternsAresAllC{} & \accuracyTokenAresAllC{}                                  & \accuracyCharactersAresAllC{} & \precisionAresAllC{} & \recallAresAllC{} \\
\indexTwoFilesD{} & \bugzillaIdTwoFilesD{} & \numberOfMembersTwoFilesD{} & \recommendationsLaseTwoFilesD{} & \foundPatternsLaseTwoFilesD{} & \accuracyTokenLaseTwoFilesD{} & \accuracyCharactersLaseTwoFilesD{} & \precisionLaseTwoFilesD{} & \recallLaseTwoFilesD{} & \recommendationsAresTwoFilesD{} & \foundPatternsAresTwoFilesD{} & \accuracyTokenAresTwoFilesD{}                                       & \accuracyCharactersAresTwoFilesD{}                                            & \precisionAresTwoFilesD{} 		& \recallAresTwoFilesD{} 		& \recommendationsAresAllD{} & \foundPatternsAresAllD{} & \accuracyTokenAresAllD{}                                  & \accuracyCharactersAresAllD{} & \precisionAresAllD{} & \recallAresAllD{} \\
\indexTwoFilesE{} & \bugzillaIdTwoFilesE{} & \numberOfMembersTwoFilesE{} & \recommendationsLaseTwoFilesE{} & \foundPatternsLaseTwoFilesE{} & \accuracyTokenLaseTwoFilesE{} & \accuracyCharactersLaseTwoFilesE{} & \precisionLaseTwoFilesE{} & \recallLaseTwoFilesE{} & \recommendationsAresTwoFilesE{} & \foundPatternsAresTwoFilesE{} & \accuracyTokenAresTwoFilesE{}                                       & \accuracyCharactersAresTwoFilesE{}                                            & \precisionAresTwoFilesE{} 		& \recallAresTwoFilesE{} 		& \recommendationsAresAllE{} & \foundPatternsAresAllE{} & \accuracyTokenAresAllE{}                                  & \accuracyCharactersAresAllE{} & \precisionAresAllE{} & \recallAresAllE{} \\
\indexTwoFilesF{} & \bugzillaIdTwoFilesF{} & \numberOfMembersTwoFilesF{} & \recommendationsLaseTwoFilesF{} & \foundPatternsLaseTwoFilesF{} & \accuracyTokenLaseTwoFilesF{} & \accuracyCharactersLaseTwoFilesF{} & \precisionLaseTwoFilesF{} & \recallLaseTwoFilesF{} & \recommendationsAresTwoFilesF{} & \foundPatternsAresTwoFilesF{} & \accuracyTokenAresTwoFilesF{}                                       & \accuracyCharactersAresTwoFilesF{}                                            & \precisionAresTwoFilesF{} 		& \recallAresTwoFilesF{} 		& \recommendationsAresAllF{} & \foundPatternsAresAllF{} & \accuracyTokenAresAllF{}                                  & \accuracyCharactersAresAllF{} & \precisionAresAllF{} & \recallAresAllF{} \\
\indexTwoFilesG{} & \bugzillaIdTwoFilesG{} & \numberOfMembersTwoFilesG{} & \recommendationsLaseTwoFilesG{} & \foundPatternsLaseTwoFilesG{} & \accuracyTokenLaseTwoFilesG{} & \accuracyCharactersLaseTwoFilesG{} & \precisionLaseTwoFilesG{} & \recallLaseTwoFilesG{} & \recommendationsAresTwoFilesG{} & \foundPatternsAresTwoFilesG{} & \accuracyTokenAresTwoFilesG{}                                       & \accuracyCharactersAresTwoFilesG{}                                            & \precisionAresTwoFilesG{} 		& \recallAresTwoFilesG{} 		& \recommendationsAresAllG{} & \foundPatternsAresAllG{} & \accuracyTokenAresAllG{}                                  & \accuracyCharactersAresAllG{} & \precisionAresAllG{} & \recallAresAllG{} \\
\indexTwoFilesH{} & \bugzillaIdTwoFilesH{} & \numberOfMembersTwoFilesH{} & \recommendationsLaseTwoFilesH{} & \foundPatternsLaseTwoFilesH{} & \accuracyTokenLaseTwoFilesH{} & \accuracyCharactersLaseTwoFilesH{} & \precisionLaseTwoFilesH{} & \recallLaseTwoFilesH{} & \recommendationsAresTwoFilesH{} & \foundPatternsAresTwoFilesH{} & \accuracyTokenAresMinTwoFilesH{}$/$\accuracyTokenAresMaxTwoFilesH{} & \accuracyCharactersAresMinTwoFilesH{}$/$\accuracyCharactersAresMaxTwoFilesH{} & \textbf{\precisionAresTwoFilesH{}} 	& \textbf{\recallAresTwoFilesH{}} 	& \recommendationsAresAllH{} & \foundPatternsAresAllH{} & \accuracyTokenAresMinAllH{}$/$\accuracyTokenAresMaxAllH{} & \accuracyCharactersAresMinAllH{}$/$\accuracyCharactersAresMaxAllH{} & \precisionAresAllH{} & \recallAresAllH{} \\
\indexTwoFilesI{} & \bugzillaIdTwoFilesI{} & \numberOfMembersTwoFilesI{} & \recommendationsLaseTwoFilesI{} & \foundPatternsLaseTwoFilesI{} & \accuracyTokenLaseTwoFilesI{} & \accuracyCharactersLaseTwoFilesI{} & \precisionLaseTwoFilesI{} & \recallLaseTwoFilesI{} & \recommendationsAresTwoFilesI{} & \foundPatternsAresTwoFilesI{} & \accuracyTokenAresTwoFilesI{}                                       & \accuracyCharactersAresTwoFilesI{}                                            & \precisionAresTwoFilesI{} 		& \recallAresTwoFilesI{} 		& \recommendationsAresAllI{} & \foundPatternsAresAllI{} & \accuracyTokenAresAllI{}                                  & \accuracyCharactersAresAllI{} & \precisionAresAllI{} & \recallAresAllI{} \\
\indexTwoFilesJ{} & \bugzillaIdTwoFilesJ{} & \numberOfMembersTwoFilesJ{} & \recommendationsLaseTwoFilesJ{} & \foundPatternsLaseTwoFilesJ{} & \accuracyTokenLaseTwoFilesJ{} & \accuracyCharactersLaseTwoFilesJ{} & \precisionLaseTwoFilesJ{} & \recallLaseTwoFilesJ{} & \recommendationsAresTwoFilesJ{} & \foundPatternsAresTwoFilesJ{} & \accuracyTokenAresTwoFilesJ{}                                       & \accuracyCharactersAresTwoFilesJ{}                                            & \textbf{\precisionAresTwoFilesJ{}} 	& \recallAresTwoFilesJ{} 		& \recommendationsAresAllJ{} & \foundPatternsAresAllJ{} & \accuracyTokenAresMinAllJ{}$/$\accuracyTokenAresMaxAllJ{} & \accuracyCharactersAresMinAllJ{}$/$\accuracyCharactersAresMaxAllJ{} & \precisionAresAllJ{} & \recallAresAllJ{} \\
\indexTwoFilesK{} & \bugzillaIdTwoFilesK{} & \numberOfMembersTwoFilesK{} & \recommendationsLaseTwoFilesK{} & \foundPatternsLaseTwoFilesK{} & \accuracyTokenLaseTwoFilesK{} & \accuracyCharactersLaseTwoFilesK{} & \precisionLaseTwoFilesK{} & \recallLaseTwoFilesK{} & \recommendationsAresTwoFilesK{} & \foundPatternsAresTwoFilesK{} & \accuracyTokenAresTwoFilesK{}                                       & \accuracyCharactersAresTwoFilesK{}                                            & \precisionAresTwoFilesK{} 		& \recallAresTwoFilesK{} 		& \recommendationsAresAllK{} & \foundPatternsAresAllK{} & \accuracyTokenAresAllK                                    & \accuracyCharactersAresAllK{} & \precisionAresAllK{} & \recallAresAllK{} \\
\indexTwoFilesL{} & \bugzillaIdTwoFilesL{} & \numberOfMembersTwoFilesL{} & \recommendationsLaseTwoFilesL{} & \foundPatternsLaseTwoFilesL{} & \accuracyTokenLaseTwoFilesL{} & \accuracyCharactersLaseTwoFilesL{} & \precisionLaseTwoFilesL{} & \recallLaseTwoFilesL{} & \recommendationsAresTwoFilesL{} & \foundPatternsAresTwoFilesL{} & \accuracyTokenAresMinTwoFilesL{}$/$\accuracyTokenAresMaxTwoFilesL{} & \accuracyCharactersAresMinTwoFilesL{}$/$\accuracyCharactersAresMaxTwoFilesL{} & \textbf{\precisionAresTwoFilesL{}} 	& \textbf{\recallAresTwoFilesL{}} 	& \recommendationsAresAllL{} & \foundPatternsAresAllL{} & \accuracyTokenAresMinAllL{}$/$\accuracyTokenAresMaxAllL{} & \accuracyCharactersAresMinAllL{}$/$\accuracyCharactersAresMaxAllL{} & \precisionAresAllL{} & \recallAresAllL{} \\
\indexTwoFilesM{} & \bugzillaIdTwoFilesM{} & \numberOfMembersTwoFilesM{} & \recommendationsLaseTwoFilesM{} & \foundPatternsLaseTwoFilesM{} & \accuracyTokenLaseTwoFilesM{} & \accuracyCharactersLaseTwoFilesM{} & \precisionLaseTwoFilesM{} & \recallLaseTwoFilesM{} & \recommendationsAresTwoFilesM{} & \foundPatternsAresTwoFilesM{} & \accuracyTokenAresTwoFilesM{}                                       & \accuracyCharactersAresTwoFilesM{}                                            & \precisionAresTwoFilesM{} 		& \recallAresTwoFilesM{} 		& \recommendationsAresAllM{} & \foundPatternsAresAllM{} & \accuracyTokenAresAllM{}                                  & \accuracyCharactersAresAllM{} & \precisionAresAllM{} & \recallAresAllM{} \\
\indexTwoFilesN{} & \bugzillaIdTwoFilesN{} & \numberOfMembersTwoFilesN{} & \recommendationsLaseTwoFilesN{} & \foundPatternsLaseTwoFilesN{} & \accuracyTokenLaseTwoFilesN{} & \accuracyCharactersLaseTwoFilesN{} & \precisionLaseTwoFilesN{} & \recallLaseTwoFilesN{} & \recommendationsAresTwoFilesN{} & \foundPatternsAresTwoFilesN{} & \accuracyTokenAresTwoFilesN{}                                       & \accuracyCharactersAresTwoFilesN{}                                            & \precisionAresTwoFilesN{} 		& \textbf{\recallAresTwoFilesN{}} 	& \recommendationsAresAllN{} & \foundPatternsAresAllN{} & \accuracyTokenAresAllN{}                                  & \accuracyCharactersAresAllN{} & \precisionAresAllN{} & \recallAresAllN{} \\
\indexTwoFilesO{} & \bugzillaIdTwoFilesO{} & \numberOfMembersTwoFilesO{} & \recommendationsLaseTwoFilesO{} & \foundPatternsLaseTwoFilesO{} & \accuracyTokenLaseTwoFilesO{} & \accuracyCharactersLaseTwoFilesO{} & \precisionLaseTwoFilesO{} & \recallLaseTwoFilesO{} & \recommendationsAresTwoFilesO{} & \foundPatternsAresTwoFilesO{} & \accuracyTokenAresTwoFilesO{}                                       & \accuracyCharactersAresTwoFilesO{}                                            & \precisionAresTwoFilesO{} 		& \textbf{\recallAresTwoFilesO{}} 	& \recommendationsAresAllO{} & \foundPatternsAresAllO{} & \accuracyTokenAresAllO{}                                  & \accuracyCharactersAresAllO{} & \precisionAresAllO{} & \recallAresAllO{} \\
\indexTwoFilesP{} & \bugzillaIdTwoFilesP{} & \numberOfMembersTwoFilesP{} & \recommendationsLaseTwoFilesP{} & \foundPatternsLaseTwoFilesP{} & \accuracyTokenLaseTwoFilesP{} & \accuracyCharactersLaseTwoFilesP{} & \precisionLaseTwoFilesP{} & \recallLaseTwoFilesP{} & \recommendationsAresTwoFilesP{} & \foundPatternsAresTwoFilesP{} & \accuracyTokenAresMinTwoFilesP{}$/$\accuracyTokenAresMaxTwoFilesP{} & \accuracyCharactersAresMinTwoFilesP{}$/$\accuracyCharactersAresMaxTwoFilesP{} & \precisionAresTwoFilesP{} 		& \textbf{\recallAresTwoFilesP{}} 	& \recommendationsAresAllP{} & \foundPatternsAresAllP{} & \accuracyTokenAresMinAllP{}$/$\accuracyTokenAresMaxAllP{} & \accuracyCharactersAresMinAllP{}$/$\accuracyCharactersAresMaxAllP{} & \precisionAresAllP{} & \recallAresAllP{} \\
\indexTwoFilesQ{} & \bugzillaIdTwoFilesQ{} & \numberOfMembersTwoFilesQ{} & \recommendationsLaseTwoFilesQ{} & \foundPatternsLaseTwoFilesQ{} & \accuracyTokenLaseTwoFilesQ{} & \accuracyCharactersLaseTwoFilesQ{} & \precisionLaseTwoFilesQ{} & \recallLaseTwoFilesQ{} & \recommendationsAresTwoFilesQ{} & \foundPatternsAresTwoFilesQ{} & \accuracyTokenAresTwoFilesQ{}                                       & \accuracyCharactersAresTwoFilesQ{}                                            & \precisionAresTwoFilesQ{} 		& \recallAresTwoFilesQ{} 		& \recommendationsAresAllQ{} & \foundPatternsAresAllQ{} & \accuracyTokenAresAllQ{}                                  & \accuracyCharactersAresAllQ{} & \precisionAresAllQ{} & \recallAresAllQ{} \\
\indexTwoFilesR{} & \bugzillaIdTwoFilesR{} & \numberOfMembersTwoFilesR{} & \recommendationsLaseTwoFilesR{} & \foundPatternsLaseTwoFilesR{} & \accuracyTokenLaseTwoFilesR{} & \accuracyCharactersLaseTwoFilesR{} & \precisionLaseTwoFilesR{} & \recallLaseTwoFilesR{} & \recommendationsAresTwoFilesR{} & \foundPatternsAresTwoFilesR{} & \accuracyTokenAresTwoFilesR{}                                       & \accuracyCharactersAresTwoFilesR{}                                            & \precisionAresTwoFilesR{} 		& \recallAresTwoFilesR{} 		& \recommendationsAresAllR{} & \foundPatternsAresAllR{} & \accuracyTokenAresAllR{}                                  & \accuracyCharactersAresAllR{} & \precisionAresAllR{} & \recallAresAllR{} \\
\indexTwoFilesS{} & \bugzillaIdTwoFilesS{} & \numberOfMembersTwoFilesS{} & \recommendationsLaseTwoFilesS{} & \foundPatternsLaseTwoFilesS{} & \accuracyTokenLaseTwoFilesS{} & \accuracyCharactersLaseTwoFilesS{} & \precisionLaseTwoFilesS{} & \recallLaseTwoFilesS{} & \recommendationsAresTwoFilesS{} & \foundPatternsAresTwoFilesS{} & \accuracyTokenAresTwoFilesS{}                                       & \accuracyCharactersAresTwoFilesS{}                                            & \precisionAresTwoFilesS{} 		& \recallAresTwoFilesS{} 		& \recommendationsAresAllS{} & \foundPatternsAresAllS{} & \accuracyTokenAresMinAllS{}$/$\accuracyTokenAresMaxAllS{} & \accuracyCharactersAresMinAllS{}$/$\accuracyCharactersAresMaxAllS{} & \precisionAresAllS{} & \recallAresAllS{} \\
\indexTwoFilesT{} & \bugzillaIdTwoFilesT{} & \numberOfMembersTwoFilesT{} & \recommendationsLaseTwoFilesT{} & \foundPatternsLaseTwoFilesT{} & \accuracyTokenLaseTwoFilesT{} & \accuracyCharactersLaseTwoFilesT{} & \precisionLaseTwoFilesT{} & \recallLaseTwoFilesT{} & \recommendationsAresTwoFilesT{} & \foundPatternsAresTwoFilesT{} & \accuracyTokenAresTwoFilesT{}                                       & \accuracyCharactersAresTwoFilesT{}                                            & \textbf{\precisionAresTwoFilesT{}} 	& \textbf{\recallAresTwoFilesT{}} 	& \recommendationsAresAllT{} & \foundPatternsAresAllT{} & \accuracyTokenAresAllT{}                                  & \accuracyCharactersAresAllT{} & \precisionAresAllT{} & \recallAresAllT{} \\
\indexTwoFilesU{} & \bugzillaIdTwoFilesU{} & \numberOfMembersTwoFilesU{} & \recommendationsLaseTwoFilesU{} & \foundPatternsLaseTwoFilesU{} & \accuracyTokenLaseTwoFilesU{} & \accuracyCharactersLaseTwoFilesU{} & \precisionLaseTwoFilesU{} & \recallLaseTwoFilesU{} & \recommendationsAresTwoFilesU{} & \foundPatternsAresTwoFilesU{} & \accuracyTokenAresTwoFilesU{}                                       & \accuracyCharactersAresTwoFilesU{}                                            & \precisionAresTwoFilesU{} 		& \recallAresTwoFilesU{} 		& \recommendationsAresAllU{} & \foundPatternsAresAllU{} & \accuracyTokenAresMinAllU{}$/$\accuracyTokenAresMaxAllU{} & \accuracyCharactersAresMinAllU{}$/$\accuracyCharactersAresMaxAllU{} & \precisionAresAllU{} & \recallAresAllU{} \\
\indexTwoFilesV{} & \bugzillaIdTwoFilesV{} & \numberOfMembersTwoFilesV{} & \recommendationsLaseTwoFilesV{} & \foundPatternsLaseTwoFilesV{} & \accuracyTokenLaseTwoFilesV{} & \accuracyCharactersLaseTwoFilesV{} & \precisionLaseTwoFilesV{} & \recallLaseTwoFilesV{} & \recommendationsAresTwoFilesV{} & \foundPatternsAresTwoFilesV{} & \accuracyTokenAresTwoFilesV{}                                       & \accuracyCharactersAresTwoFilesV{}                                            & \precisionAresTwoFilesV{} 		& \recallAresTwoFilesV{} 		& \recommendationsAresAllV{} & \foundPatternsAresAllV{} & \accuracyTokenAresAllV{}                                  & \accuracyCharactersAresAllV{} & \precisionAresAllV{} & \recallAresAllV{} \\
\indexTwoFilesW{} & \bugzillaIdTwoFilesW{} & \numberOfMembersTwoFilesW{} & \recommendationsLaseTwoFilesW{} & \foundPatternsLaseTwoFilesW{} & \accuracyTokenLaseTwoFilesW{} & \accuracyCharactersLaseTwoFilesW{} & \precisionLaseTwoFilesW{} & \recallLaseTwoFilesW{} & \recommendationsAresTwoFilesW{} & \foundPatternsAresTwoFilesW{} & \accuracyTokenAresMinTwoFilesW{}$/$\accuracyTokenAresMaxTwoFilesW{} & \accuracyCharactersAresMinTwoFilesW{}$/$\accuracyCharactersAresMaxTwoFilesW{} & \textbf{\precisionAresTwoFilesW{}} 	& \textbf{\recallAresTwoFilesW{}} 	& \recommendationsAresAllW{} & \foundPatternsAresAllW{} & \accuracyTokenAresMinAllW{}$/$\accuracyTokenAresMaxAllW{} & \accuracyCharactersAresMinAllW{}$/$\accuracyCharactersAresMaxAllW{} & \precisionAresAllW{} & \recallAresAllW{} \\\hline
 Avg.  &  & \numberOfMembersTwoFilesAvg{} & \recommendationsLaseTwoFilesAvg{} & \foundPatternsLaseTwoFilesAvg{} & \accuracyTokenLaseTwoFilesAvg{} & \accuracyCharactersLaseTwoFilesAvg{} & \precisionLaseTwoFilesAvg{} & \recallLaseTwoFilesAvg{} & \recommendationsAresTwoFilesAvg{} & \foundPatternsAresTwoFilesAvg{} & \accuracyTokenAresMinTwoFilesAvg{}$/$\accuracyTokenAresMaxTwoFilesAvg{} & \accuracyCharactersAresMinTwoFilesAvg{}$/$\accuracyCharactersAresMaxTwoFilesAvg{} & \precisionAresTwoFilesAvg{} & \recallAresTwoFilesAvg{} & \recommendationsAresAllAvg{} & \foundPatternsAresAllAvg{} & \accuracyTokenAresMinAllAvg{}$/$\accuracyTokenAresMaxAllAvg{} & \accuracyCharactersAresMinAllAvg{}$/$\accuracyCharactersAresMaxAllAvg{} & \precisionAresAllAvg{} & \recallAresAllAvg{} \\
	\end{tabular} 
    \label{fig:eval_lase_groups}
    \begin{center}
     \scriptsize{$m$: Available Locations; $\triangle$: Generated Recommendations; $\checkmark$ Correct Recommendations;\\\textit{A$_T$\%}: Token Accuracy; \textit{A$_T$\%}: Character Accuracy; \textit{P\%} : Precision; \textit{R\%} : Recall;}

    \end{center}

\end{table*}

\section{Evaluation}
\label{sec:evaluation}
We first evaluate the accuracy of ARES and compare it with that of LASE~\cite{MENG2013},
before we show that the improved accuracy does not gravely impact
precision, recall, or execution times.
For all measurements we use a workstation with 128 GB of RAM and a
3.6 GHz Intel Xeon CPU, OpenJDK 8 and Ubuntu 16.10.
We also discuss the current limitations of ARES in this section.

\subsection{Accuracy}
\label{sec:evaluation_accuracy}
Our benchmarks comprise several real-world source code archives that vary with respect to the
number and size of the sets of training examples (similar code changes). We always check how close
the recommended code changes get to changes that can actually be found in the commits. 

\textbf{Eclipse---Two Input Code Changes.}
In this part, we use 2 groups of similar code changes (Bugzilla Ids 77644, 82429)
from the \textit{Eclipse JDT} project~\cite{ECLIPSEJDT} and 21 groups
from the \textit{Eclipse SWT} project~\cite{ECLIPSESWT}. Meng et al.~\cite{MENG2013} used the same \textit{manually}
collected 23 groups for their
evaluation of LASE. All these code changes are present in commits of the respective repositories.

Table~\ref{fig:eval_lase_groups}
lists the \textit{Bugzilla Ids} of the changes; \textit{m} is the size of the group
of similar code changes that exist in the repositories for a bug.
In most cases, \textit{m} is equal to the evaluation provided by Meng et al. The 
only exceptions are the rows with Ids 20 and 21 for which we identified
a different number of changes in the repository.

For the next two segments of Table~\ref{fig:eval_lase_groups} (\textit{LASE---Two Input Code Changes}, \textit{ARES---Two Input Code Changes})
we use the same \textit{two} changes of a group for the creation of the patterns.
We carefully selected the two changes to reproduce the same precision and recall values that Meng et al.~\cite{MENG2013} list in
their evaluation.

For each of the 23 bug fixes, the \boldmath{$\triangle$}-columns list
the number of code locations for which LASE and ARES create a recommendation.
The columns marked with \boldmath{$\checkmark$} show how many of
the \textit{m} manually identified locations the tools find.
For each of them column \boldmath{$A_T$} and \boldmath{$A_C$} give the
accuracy, i.e., the closeness of the recommendation
to the code that the original developers of the change 
wrote. Thus, a perfect accuracy means that the 
recommendation has the same statements, moved code and identifiers
that are present in the repository. For developers this is
important because a higher accuracy means less work to actually apply
a recommendation to a project. 
To measure the accuracy we compute the Levenshtein distance~\cite{levenshtein} (LD)
between the method body of the recommendation and the 
method body of the changed code from the repository.
We use two variants. The first uses AST tokens (\textbf{T}), the second actual characters (\textbf{C}).
A high token accuracy \boldmath{$A_T$} shows that the recommendation is
accurate with respect to the syntax (e.g., uses the same number of \textit{if} statements).
We use \boldmath{$A_T$} = $1 - {LD_T}/{max(|r_T|, |m_T|)}$, where $|r_T|$ is the
number of tokens in the recommendation and $|m_T|$ is the number of tokens in the method body from the 
repository. As comments are not part of any token, they are ignored for this measurement.
We also define \boldmath{$A_C$} = $1 - {LD_C}/{max(|r_C|, |m_C|)}$, where $|r_C|$ is the
number of characters in the recommendation and $|m_C|$ is the number of characters in the method body from the 
repository. This includes both comments and whitespace. Each cell in the accuracy columns contains the mean of the correctly
identified recommendations (\boldmath{$\checkmark$}). 

Across all groups of code changes, ARES can produce more accurate recommendations compared to LASE
and even achieves a perfect accuracy (100\%) for 11 of the 23 groups. For these groups, the
recommendations are identical to the changes that were actually performed by a human developer. A
closer inspection of the recommendations by LASE shows that LASE ignores code transformations that
occur only in one of the input examples which reduces \boldmath{$A_T$}. For some cases, the ARES
accuracy column contains two values. In these cases, the pattern contains a choice annotation and
ARES generates several recommendation variants. \boldmath{$A_T$} and \boldmath{$A_C$} give the
minimal and the maximal accuracy values for the corresponding groups. In 3 of the 5 cases (Ids: 2,
12, 23), even the minimal values for \boldmath{$A_T$} of ARES outperform LASE. Often (for 11 groups)
ARES achieves perfect accuracy values (100\%) in \boldmath{$A_C$}. Hence,
ARES recommends exactly the code a developer has written, including comments, coding style, and whitespace.

\textbf{Eclipse---All Code Changes.}
In some cases (e.g., patterns for critical bugs) a high recall is more important
than a high precision (and accuracy). For these cases, ARES supports more
than two input changes.
Each additional change in the training set
can increase the number of wildcards and hence
the recall. However, each additional wildcard can also
decrease precision and accuracy. 
To examine the effects of additional input changes
we added a third set of evaluation results in Table~\ref{fig:eval_lase_groups} (\textit{ARES---All Code Changes}).
For these results, ARES uses all input changes to create a pattern.
Thus, we maximize the recall and minimize precision and accuracy for each of the groups.

For this set, the accuracy of ARES is still at 94\% (87\% in the minimal choice case)
on average.
In most cases, the accuracy of ARES in this configuration is also higher than
the accuracy of LASE with only two input changes. This means that even with
the most general pattern for a group and thus the pattern with the least 
accuracy, ARES still achieves higher accuracy values than LASE. Also, ARES still has a perfect
accuracy for 8 groups.

\textbf{JUnit---All Code Changes.}
The above analysis shows that ARES achieves a higher accuracy for the 23
groups of code changes from Eclipse. A threat to the validity of these results is that
it is possible to optimize a system for such a small dataset. To examine whether ARES also has higher 
accuracy values for a larger dataset,
we use 3,904 groups of code changes from JUnit~\cite{JUNIT} taken from the results of
\cthree{}~\cite{Kreutzer2016}.
\cthree{} is a tool that identifies groups of similar code changes in code repositories
with the help of clustering algorithms.
The groups by \cthree{} are suitable inputs for ARES and LASE.
This time we use all input examples (not only two) to train
ARES and LASE. 

Whereas it is again possible to measure the accuracy (as the manual code
changes are in the repository) there cannot be numbers for precision
and recall as \textit{m} is unknown for this dataset.

Table~\ref{table:resultsJunit} shows the accuracy results for JUnit.
For 482 sets of input changes both LASE and ARES produce recommendations
and generate a recommendation for the input changes. We excluded input changes 
for which one of the tools 
produces an error or does not recommend a change.
Similar to the evaluation above, ARES reaches a high
accuracy of 91\% to 100\%. Again for most
groups, the accuracy of ARES is higher than that achieved by LASE. This is also
true for the minimal accuracy values.

\begin{table}[b]
	
	\caption{Accuracy on JUnit. \label{table:resultsJunit}}
  \vspace*{-0.3cm}
	\scriptsize
  \centering
	\begin{tabular}{l|r|r}
	                        & \hspace{0.4cm} LASE \hspace{0.4cm}              & ARES (Min/Max)      \\\hline
	Groups of Code Changes                  &   \multicolumn{2}{c}{3,904}\\
	Shared Recommendations &   \multicolumn{2}{c}{482}\\
	Shared Recommendations \boldmath{$A_T$} (Mean) \%&     90     &  91/97    \\
	Shared Recommendations \boldmath{$A_T$} (Median) \%&     100     &  100/100    \\
	Shared Recommendations \boldmath{$A_C$} (Mean) \%&     76     &  90/95   \\
	Shared Recommendations \boldmath{$A_C$} (Median) \%&     82     &  100/100    \\
	\end{tabular}
\end{table}

\subsection{Precision, Recall, and Time Measurements}
\textbf{Precision \& Recall.}
Let us now compare the precision \textit{P} ($\frac{\checkmark}{\triangle}$),
and recall \textit{R} ($\frac{\checkmark}{m}$) values. The higher the precision and recall, the better is the tool.
Overall, the precision of ARES is similar to that of LASE.
For 20 of the 23 groups of code changes the precision of ARES is identical to or higher than that of LASE.
The average recall of ARES is still at a high level (76\%) but below that of LASE (87\%).
However, Christakis and Bird~\cite{CHRISTAKIS2016} found in their study that the recall
is less important to developers compared to a high precision. Hence, with respect to 
precision and recall there are (almost) no costs for the improved accuracy of ARES.

A closer look reveals that ARES is more precise than LASE for three change groups (Ids: 10, 12,
23) because of two reasons. First, ARES can handle variations in the input changes with
appropriate wildcards. Instead, LASE uses the \textit{Maximum Common Embedded Subtree Extraction}
algorithm to identify the common AST that all training examples share and which is more general in
some cases. Second, ARES keeps common code, even if it is unrelated to the changes, whereas LASE
identifies code that has no dependencies to the code change and excludes it from the pattern.

While it can be beneficial to keep some unrelated code in the pattern,
keeping too much of it causes a loss of precision (Id 8) or recall. There is room for more
research concerning this issue. In contrast to LASE, ARES is also currently limited to the
method body and does not include the method signature in the patterns.
This lack of a signature in the pattern lowers the precision in two cases (Ids: 2, 20).

The groups 20 and 23 are outliers as the precision of both ARES and LASE is considerably lower
compared to the other groups. The reason is that the input examples for both groups only add the
same statements and otherwise have very little code in common. This leads to short and very
general patterns.

\begin{figure}[t]
 \scriptsize
 \begin{center}
 \begin{tikzpicture}
  \begin{axis}
    [
        yticklabel style={anchor=west, xshift=-2.8cm},
		ytick={1,2,3,4},
    ytick style={draw=none},
		yticklabels={ARES --- \textit{Pattern Use}, LASE --- \textit{Pattern Use}, ARES --- \textit{Pattern Creation}, LASE --- \textit{Pattern Creation}},
    width=0.8\linewidth,
    height=4.6cm,
    axis lines=left,
		xmin=-0.01,
    xlabel={Execution time (sec.)},
    xlabel style={yshift=0.5em},
    xmax=80,
    ymax=4.6,
    ymin=0.4,
    y axis line style=-
    ]
               
                                \addplot+[black,solid,mark=o, mark options={black,solid, scale=0.8},
                boxplot prepared={
			median=19.169849,
                        upper quartile=41.511050,
                        lower quartile=14.900138,
                        upper whisker=79.250374,
                        lower whisker=3.584652
                },
                ] coordinates {};
                
                              \addplot+[black,solid,mark=o, mark options={black,solid, scale=0.8},
                boxplot prepared={
 			median=5.150711,
                        upper quartile=7.769817,
                        lower quartile=4.577096,
                        upper whisker=13.907895,
                        lower whisker=0.331092
               },
                ] coordinates {};

                                \addplot+[black,solid,mark=o, mark options={black,solid, scale=0.8},
                boxplot prepared={
			median=0.054162,
                        upper quartile=0.681364,
                        lower quartile=0.016469,
                        upper whisker=4.163283,
                        lower whisker=0.010198
                },
                ] coordinates {};

                              \addplot+[black,solid,mark=o, mark options={black,solid, scale=0.8},
                boxplot prepared={
 			median=0.109872,
                        upper quartile=1.160704,
                        lower quartile=0.029010,
                        upper whisker=13.292958,
                        lower whisker=0.011363
               },
                ] coordinates {};

  \end{axis}
\end{tikzpicture}
\end{center}
\vspace*{-0.5cm}
\caption{Time per change group for two input code changes (25\%$/$75\% quartiles,
whiskers: minimum/maximum).}
\label{fig:time_boxplot_create}
\vspace*{-0.3cm}
\end{figure}
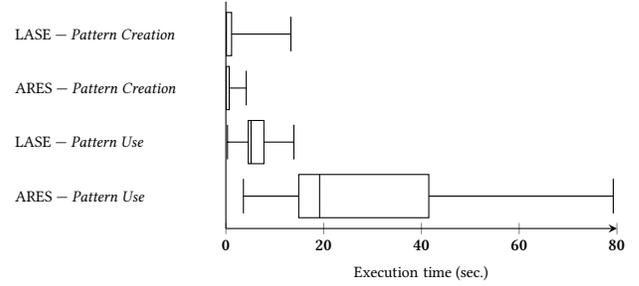

\textbf{Time Measurements.}
Here we compare the times that ARES and LASE take to create the patterns and
to use them when they browse the projects 
for possible pattern applications (Eclipse SWT: $\sim$2,000 Java files, $\sim$375,000 lines of code;
Eclipse JDT: $\sim$5,750 Java files, $\sim$1,372,000 LOC).
Fig.~\ref{fig:time_boxplot_create} shows the measurements.
Each code change (row in Table~\ref{fig:eval_lase_groups}) corresponds to
one measurement point. The lines in the boxes are the medians of the time measurements.
The boxes of the plot define 
the 25\% and 75\% quartiles, the whiskers show the minimum and maximum.

ARES creates patterns faster than LASE (\textit{Pattern Creation} in
Fig.~\ref{fig:time_boxplot_create}) because the ChangeDistiller~\cite{FLURI2007} tree differencing
algorithm that LASE uses has a higher runtime than MTDIFF~\cite{MTDIFF}. However, the search for
pattern applications in ARES (\textit{Pattern Use} in Fig.~\ref{fig:time_boxplot_create}) is
slower because the patterns of ARES are currently limited to method bodies and do not contain
method signatures. LASE can use the method signature to filter out methods and thus has to
inspect fewer method bodies. Still, even for such large repositories ARES completes the search for
one pattern within a minute in most cases.

\subsection{Limitations and Threats to Validity}
The major limitation of ARES is the current restriction to method bodies.
Due to this restriction we also had to exclude the group of code changes with
Bugzilla Id 74139 that Meng et al.~\cite{MENG2013} use
in their evaluation of LASE. This was necessary as the code changes in this group
only have a common signature but no common statements.

Whereas both the search and recommendation creation of ARES
already support larger changes, extending the pattern creation to
work with full classes would need annotations that, for example,
define when methods or fields can be reordered. 
The pattern creation would also require
new rules and transformations for wildcards outside of code blocks.

Another limitation is that ARES (like LASE) can only search for one pattern
at a time.
Similar to their application in code clone detection tools,
suffix trees~\cite{KAMIYA2002} may accelerate the search for several patterns in
parallel.

The threat to validity of the evaluation is the implementation of LASE that we
built from a publicly available version~\cite{LASEURL}. We had
to apply several bug fixes to be able to replicate the evaluation results of 
Meng et al.~\cite{MENG2013}. Despite our efforts, there are still some
small differences left. We also changed
the implementation to bypass the UI to enable script-based
performance measurements. We argue that if we have introduced new errors
they are probably small since we obtained most of the original results.

\section{Related Work}
\label{sec:related_work}
The book \textit{Recommendation Systems in Software Engineering}~\cite{ROBILLIARD2010}
and the study by Gasparic and Janes~\cite{GASPARIC2016} provide an overview of
the related work. We discuss the works closest to ARES.

\textbf{Example-based recommendation.} LASE~\cite{MENG2013} is closest to ARES.
The main differences are that LASE
does not handle moved code parts accurately and thus
the accuracy of the recommendations is lower.
RASE~\cite{MENG2015} relies on LASE to create a generalized pattern from examples.
With this pattern it refactors the code
without altering its semantics to replace all changed locations with a single
unifying code fragment. Thus, it has the same low accuracy as LASE and also only supports
refactorings.
In contrast to ARES, SYDIT~\cite{MENG2011} generates an edit script from
just one example which limits its generalization ability. Developers also
must find pattern applications manually
with SYDIT whereas ARES finds them automatically.
Critics~\cite{ZHANG2015} addresses the review process. The
developer provides a generalization of a single code change as input. Critics
then finds matching spots in the code where a similar change may have been
forgotten or where an inconsistent modification may have occurred.
For Critics, it is the task of the developer to provide the generalization and
to take care of the design of the pattern to increase the accuracy.
ARES creates the pattern automatically.

REFAZER~\cite{REFAZER} is another tool that learns code transformations from examples.
It uses a domain specific language (DSL) to represent code change patterns. In contrast to ARES,
the algorithms that generate this pattern do not support code movements and thus
if a code pattern relies on movements it cannot be as accurate as ARES.
Currently, a direct comparison is not possible
as REFAZER only supports C\# and Python whereas ARES uses Java.

Santos et al.~\cite{SANTOS2017} compare three different ways (structural, AST-based,
information retrieval) to
search for additional code change locations based on up to two examples.
If a search finds such a location, their system applies a set of code changes.
If this is successful, their system recommends the code change.
Their AST-based approach uses the longest common subsequence (LCS)~\cite{LCS} and 
thus is less precise then ARES in regard to moved code. Also this approach
is time consuming as it requires the execution of the LCS on many different ASTs.
ARES is faster as its backtracking algorithm can abort the search for possible
pattern applications earlier.
Their information retrieval approach relies on the similarity of code parts (e.g., variable names)
and is less precise then ARES that includes the code structure in the pattern.
Their structural approach searches for methods with similar signatures,
packages, etc. The authors argue that this is useful in combination with
the AST-based and information retrieval approaches.
It is possible to include their structural approach in ARES to increase precision.

Tate et al.~\cite{TATE2010_learn_opt_by_proof} use a proof-checker to learn transformations from examples.
They can only find provably semantic-preserving transformations.

Padioleau et al.~\cite{PADIOLEAU2008} introduce semantic patches to
manually generalize patches obtained from standard diff tools.
Andersen et al.~\cite{ANDERSEN2008, ANDERSEN2012} extend this idea and introduce
a tool that creates generalized (semantic) patches from hand-picked diffs. 
ARES is more accurate as it supports code movements whereas diffs are limited
to \textit{insert} and \textit{delete} operations.

\textbf{Code completion.} 
Cookbook~\cite{JACOBELLIS14} uses generalizations from examples to suggest code
completions while the developer is typing. In contrast to ARES it is a line based
approach and does not support code movements and thus has a lower accuracy if
code movements are necessary.
MAPO~\cite{MAPO2009} and Precise~\cite{ZHANG2012_PRECISE} 
search for similar API usages in code repositories.
MAPO recommends frequently
used call sequences. Precise extracts API calls including their argument values and corresponding declarations from repositories, extracts groups 
with similar arguments using a k-nearest neighbor algorithm, and generates API usage recommendations, including possible parameters.
Bruch et al.~\cite{BRUCH2009} also use a k-nearest neighbor algorithm and find code fragments that are similar
to code that is currently being developed. Bajracharya et al.~\cite{BAJRACHARYA2010} use
structural semantic indexing for this purpose.
All focus on code completions and thus small coding suggestions. In contrast, ARES strives to
suggest larger transformations including changes of complete methods.

\textbf{Specialized recommendation.} In addition to example-based approaches, there are also tools for specific tasks.
CFix~\cite{JIN2012_CFIX} uses predefined patterns to automatically fix concurrency bugs.
Weimer et al.~\cite{WEIMER2009} and AutoFix-E~\cite{WEI2010_AUTOFIX_E} generate bug fixes automatically,
but are limited to
testable fixes and specifications.
Robbes and Lanza~\cite{ROBBES2008_example_prog_trans} learn
code transformations by examining edit operations in the IDE.
Their tool cannot automatically find locations. 
Stratego/XT~\cite{VISSER2001_STRATEGO} and DMS~\cite{BAXTER2004_DMS} offer DSLs to 
specify AST transformations.
Manniesing et al.~\cite{MANNIESING2000_automaticsimd} focus on loop transformations.
Other works solely 
locate patterns in source code, e.g., the
Dependency Query Language~\cite{WANG2010}
or the Program Query Language~\cite{MARTIN2005}.
In contrast to ARES, developers must specify patterns manually. They also do not get recommendations. 
Miller et al.~\cite{MILLER2001} let developers select multiple code fragments and
then they apply a change to all of them in the same fashion.
In contrast to ARES this is limited to identical changes and developers have to find the code locations themselves.
Thung et al.~\cite{THUNG2016} introduce a recommendation system that supports developers
in backporting Linux drivers. Their framework is specifically tailored to the backporting and is not intended for general 
code change recommendations.

\section{Conclusion}
We presented the novel Accurate REcommendation System (ARES) that specializes on code movements to
increase the accuracy of code recommendations. A higher accuracy means that ARES generates code
recommendations that better reflect what a developer would have written. ARES achieves the higher
accuracy results with a pattern design that expresses code movements more accurately compared to
the state-of-the-art. Similar to other tools, ARES generates these patterns from source code
training examples. The generated patterns of ARES contain only plain Java code with a set of
annotations and therefore are not a black box for developers. Thus, developers can read and
manually adapt the patterns.

We also presented in detail how ARES generates, generalizes, and applies these patterns. With
these techniques ARES achieves an average recommendation accuracy of 96\% in our evaluation and
outperforms LASE. Precision and recall are on par. The execution time is a bit higher, but still
below two minutes for large real-world source code archives.

For reproducibility and to kindle further research, we open-source ARES, the rules for the edit script
adjustment, all the evaluation inputs and results, including the human-readable patterns generated
by ARES (\url{https://github.com/FAU-Inf2/ARES}). 

\bibliographystyle{ACM-Reference-Format}
\bibliography{refs} 
\balance

\end{document}